\documentclass[11pt,a4paper]{article}
\usepackage{epic,eepic,amsmath,latexsym,fullpage,amssymb,color,amsthm}
\usepackage{ifthen,graphics,epsfig,xspace}
\usepackage[english]{babel} 
\usepackage{times}
\usepackage[left=.9in, right=1.in, top=.9in, bottom=.9in]{geometry}
\newlength {\squarewidth}

\newtheorem{definition}{Definition}
\newtheorem{theorem}{Theorem}
\newtheorem{lemma}{Lemma}

\newcommand{\toto}{xxx}
\newenvironment{proofT}{\noindent{\bf Proof }}
{\hspace*{\fill}$\Box_{Theorem~\ref{\toto}}$\par\vspace{3mm}}
\newenvironment{proofL}{\noindent{\bf Proof }}
{\hspace*{\fill}$\Box_{Lemma~\ref{\toto}}$\par\vspace{3mm}}

\newenvironment{lemma-repeat}[1]{\begin{trivlist}
\item[\hspace{\labelsep}{\bf\noindent Lemma~\ref{#1} }]}%
{\end{trivlist}}

\newenvironment{theorem-repeat}[1]{\begin{trivlist}
\item[\hspace{\labelsep}{\bf\noindent Theorem~\ref{#1} }]}%
{\end{trivlist}}

\newcounter{linecounter}
\newcommand{\linenumbering}{\ifthenelse{\value{linecounter}<10}
{(0\arabic{linecounter})}{(\arabic{linecounter})}}
\renewcommand{\line}[1]{\refstepcounter{linecounter}\label{#1}\linenumbering}
\newcommand{\resetline}{\setcounter{linecounter}{0}}
\renewcommand{\thelinecounter}{\ifnum \value{linecounter} > 
9\else 0\fi \arabic{linecounter}}
\newenvironment{mycenter}{\begin{list}{}{%
\topsep0pt \partopsep0pt \parskip0pt \leftmargin0pt}%
\centering\item\relax}{\end{list}}
\newsavebox{\mybox}
\newenvironment{myalgo}[1][aaaaa\=aa\=aa\=aa\=aa\=\kill]{
\begin{lrbox}{\mybox}
\footnotesize

\resetline
\begin{minipage}[t]{0pt}
\begin{tabbing}#1
}{\end{tabbing}
\end{minipage}\normalsize
\end{lrbox}
\begin{mycenter}
\fbox{\usebox{\mybox}}
\end{mycenter}
}

\newcommand{\REG}{\mathit{REG}}
\newcommand{\Omit}[1]{}

\newlength{\reduceunderfig}
\renewcommand{\paragraph}[1]{\vspace{0.18cm}\noindent \textbf{#1}~~}
\let\bibitemOld\bibitem
\renewcommand{\bibitem}[1]{\vspace{-0.2cm} \bibitemOld{#1}}
\let\subsectionOld\subsection
\renewcommand{\subsection}[1]{\vspace{-0.22cm}\subsectionOld{#1}\vspace{-0.25cm}}
\let\subsubsectionOld\subsubsection
\renewcommand{\subsubsection}[1]{\vspace{-0.3cm}\subsubsectionOld{#1}\vspace{-0.15cm}}

\begin{document}

\title{\bf Stabilizing Server-Based Storage\\
in Byzantine Asynchronous Message-Passing Systems\\
\large{Preliminary Version}}

\author{Silvia Bonomi$^{\#}$,~~ 
Shlomi Dolev$^{\S}$,~~
Maria Potop-Butucaru$^{\dag}$,~~
Michel Raynal$^{\star,\ddag}$ ~\\~\\
$^{\#}$      MIDLAB, Universit\`a La Sapienza, Roma, Italia\\
$^{\S}$      Department of Computer Science,
            Ben-Gurion University, Beer-Sheva, Isra\"{e}l\\
$^{\dag}$    LIP6, Universit\'e P. \& M. Curie, Paris, France\\
$^{\star}$   Institut Universitaire de France\\
$^{\ddag}$   IRISA, Universit\'e de Rennes, 35042 Rennes, France \\
{\footnotesize {\tt bonomi@dis.uniroma1.it
                    maria.potop-butucaru@lip6.fr
                    dolev@cs.bgu.ac.il
                    raynal@irisa.fr}}
}

\date{}
\maketitle

\begin{abstract}
A stabilizing Byzantine single-writer single-reader (SWSR) regular 
register, which stabilizes after the first invoked write operation, 
is first presented. 
Then, new/old ordering inversions are eliminated by the use of a (bounded) 
sequence number for writes, obtaining a practically stabilizing 
SWSR atomic register. 
A practically stabilizing Byzantine single-writer multi-reader (SWMR) 
atomic register is then obtained by using several copies of SWSR atomic
 registers. Finally,  bounded time-stamps, 
with a time-stamp per writer, together  with SWMR atomic registers,
are used to construct a practically stabilizing Byzantine  multi-writer
multi-reader (MWMR) atomic  register. 
In a system of $n$ servers implementing an atomic  register,  
and in addition to transient failures, 
the constructions tolerate $t<n/8$ Byzantine servers if  communication is  
asynchronous, and  $t<n/3 $ Byzantine servers if it is  synchronous. 
The noteworthy feature of the proposed algorithms is that (to our knowledge) 
these are the first that build an atomic read/write storage 
on top of asynchronous servers prone to  transient failures,
and where up to $t$ of them can be  Byzantine.

~\\~\\
{\bf Keywords}
Asynchronous message-passing system,
Atomic read/write register, 
Byzantine server, 
Clients/servers architecture, 
Distributed algorithm, 
Fault-tolerance, 
Regular  read/write register, 
Self-stabilization,
Transient failures. 
\end{abstract}




\section{Introduction}
\paragraph{Byzantine processes and self-stabilization}
Algorithms that tolerate Byzantine faults are of extreme interest, as they 
can tolerate malicious takeovers of portions of the system,
and still achieve the desired goal. Moreover, as the program executed by 
several of the participants may include programming mistakes, it is possible
that these participants will (unintentionally) behave in a malicious way. 
Obviously, when all participants exhibit Byzantine arbitrary behavior,
the system output will be arbitrary too.

Usually,  lower bounds on the number of Byzantine participants are used  
as part of the algorithm design assumptions. 
The cases in which the lower bound is not respected are
not considered, as the system can reach an arbitrary configuration 
due to the possibly overwhelming malicious actions. 
Assume that some of the  Byzantine participants regain consistency 
(possibly by rebooting, running anti-virus software, environment change) 
so that the assumed threshold on the number of Byzantine participants is 
now respected. Will the system regain consistency, from this arbitrary 
configuration? Or in other words will the system stabilize to a correct 
behavior?

\paragraph{Related work and aim of the paper}
An active research area concerns the construction of a Byzantine-tolerant 
disk storage (e.g.,\cite{AB06,CM05,MA04} to cite a few). Many
of these papers consider registers built on top of duplicated
disks (servers), which are accessed by clients, and where
disks and clients may exhibit different type of failures.
The construction of a reliable shared memory on top of a  Byzantine 
message-passing system is addressed in~\cite{IRRS14}.

Recently, several works investigated stabilizing Byzantine algorithms
e.g.,~\cite{BPT15,DL+13,DW04}. The first of these papers is the most related 
to our research, as it constructs a stabilizing Byzantine multi-writer 
multi-reader regular register, where $t$ out of $n$ servers
(with $n\ge5t+1$) can be Byzantine. 
Such a construction relies on the {\it write operation quiescence} 
assumption, i.e., it is assumed that, after a burst of write operations 
executed by the 
writer, there exists a sufficiently long period during which the writer 
does not invoke the write operation.  Differently, we construct a 
{\it practically stabilizing Byzantine multi-writer multi-reader atomic} 
register in a client/server system 
which is able to tolerate transient failures and up to $t$ Byzantine servers.
Given $t$, our solutions require  $n\geq 8t+1$
servers when  client/server communication is asynchronous,  
and only $n\geq 3t+1$ servers when it is synchronous. 
This gap comes from the fact that, as they provide bounds on message 
transfer delays, synchronous settings allows readers and writers to use timers. 
As far as we known, our construction is the first that builds a distributed
atomic read/write memory on top of asynchronous servers, which communicate by 
message-passing, can suffer transient failures, and where some of them can 
 exhibit a Byzantine behavior.  

\paragraph{Roadmap}
The paper is made up of~\ref{sec:conclusion} sections. 
The computing model and the problem which is addressed are presented in 
Section~\ref{sec:model-and-pb}. Then, Section~\ref{sec:regular-algorithm}
presents and proves correct an algorithm that builds a stabilizing 
single-writer single-reader (SWSR) regular register. This algorithm is extended 
in Section~\ref{sec:practically-atomic} to obtain an SWSR atomic register,  
and Section~\ref{sec:single-to-multi} shows how to go from  
``single-reader'' to ``multi-reader'' and from ``single-writer'' to 
``multi-writer''. Finally Section~\ref{sec:conclusion} concludes the paper.  
Due to page limitation, the synchronous communication case
and proofs can be found in appendices.

\section{Computing Model and the Problem we Want to Solve}
\label{sec:model-and-pb}

\subsection{Computing model}
\label{sec:model}
\paragraph{Basic system model}
The basic system model we consider consists of $(n+2)$ asynchronous sequential 
processes. One of them is called ``writer'' (denoted $p_w$), another 
is called ``reader'' (denoted $p_r$), while the $n$ others are called 
``servers'' (denoted $s_1$, ..., $s_n$). 

From a communication point of view,  there are $4n$  directed   
asynchronous communication links, connecting  each server to $p_w$ and $p_r$
(one in each direction). Each link is FIFO and reliable (neither loss, 
corruption, duplication,  nor creation of messages). 

It is assumed that processing times are negligible, and are consequently 
assumed to take zero time. Only message transfers takes time.

This basic model will be later enriched in two directions: one concerning 
client processes to have $m$ reader/writer processes, and a second 
concerning the synchrony of the communication links.

\paragraph{Failure model}
At most $t<n/8$ servers can commit Byzantine failures\footnote{Actually, 
Byzantine failures can be ``mobile''~\cite{SW89,SW07}.This means that, if, 
after some time,
a server that committed Byzantine failures, starts behaving correctly, 
a server that was previously behaving correctly can become Byzantine. 
This ``failure mobility'' can occur at any time during  the periods where
there is no pending read or write operation, issued by $p_w$ or $p_r$.
In fact, in any case, the system is guaranteed to converge to exhibit the 
desired behavior once the assumptions concerning the system hold again
for a ``long enough'' period of time.}. 
Let us remember that a server commits a Byzantine failure when it behaves 
arbitrarily~\cite{LSP82}. Classical examples of a Byzantine behavior consists 
in sending erroneous values, not sending a message when this should be done, 
stopping its execution, etc.

In addition to the possibility of Byzantine servers, the local variables 
of any process (writer, reader, servers) can suffer transient failures. 
This means that their values can be arbitrarily modified~\cite{D00}. 
It is nevertheless assumed that there is a finite time $\tau_{no\_tr}$ 
(which remains always 
unknown to the processes) after which there are no more transient 
failures\footnote{This assumption is required to ensure that, despite 
asynchrony and Byzantine behaviors, the problem we are interested in can 
be solved. In fact, if the time between two successive transient faults
is long enough, the system converges and produces useful outputs 
between transient failures.}. 

From a terminology point of view, a server is {\it correct} if it does not 
commit Byzantine failures.  Hence, as the reader and the 
writer, any correct server can  suffer transient failures.

\paragraph{Configurations and executions}
Each process  (writer, reader, or server) is a state machine, 
enriched with the operations send and receive. Its  state (called 
``local state'') is defined by the current values of its local variables. 
The state of a directed  link consists of the messages that have been sent on 
this  link, and are not yet received. 

A configuration (or global state) is composed of the local state of 
each process  and the state of each link. 
Due to the ``transient failures'' behavioral assumption, 
the initial configuration can be arbitrary.

\paragraph{Underlying ss-broadcast abstraction}
It is assumed that the system has a built-in  communication abstraction, 
denoted ss-broadcast, that provides the reader and the writer
with an  operation denoted ${\sf ss\_broadcast}()$,  and each server with 
a matching  operation denoted ${\sf ss\_deliver}()$. 
When the reader or the writer (resp., server)   uses this broadcast 
abstraction, we consequently say that it ``ss-broadcasts'' (resp.,
``ss-delivers'') a message. 
This communication abstraction  is defined by the following properties. 
\begin{itemize}
\vspace{-0.2cm}
\item \emph{Termination}. If  the reader or the writer  invoke  
${\sf ss\_broadcast}(m)$ then such invocation terminates. 
\vspace{-0.2cm}
\item \emph{Eventual delivery}. 
If the reader or the writer  invokes
${\sf ss\_broadcast}(m)$ then every correct server eventually ss-delivers $m$. 
\vspace{-0.2cm}
\item \emph{Synchronized delivery}. 
If a process $p_x$ (reader or writer) invokes ${\sf ss\_broadcast}(m)$  
at time $\tau_1^x$ and returns from this invocation at time $\tau_2^x$, 
then there exists a set $S$ of $(n-2t)$  correct servers, such that, for 
each $s_i\in S$, there exists a time $\tau(i)$ such that 
$\tau_1^x<\tau(i)<\tau_2^x$ at which $s_i$ executed  ${\sf ss\_delivery}(m)$. 
\vspace{-0.2cm}
\item \emph{No duplication}. 
An invocation of ${\sf ss\_broadcast}(m)$ by a process $p$ (reader or writer) 
results in at  most one ${\sf ss\_deliver}(m)$ at any correct server $s_i$. 
\vspace{-0.2cm}
\item \emph{Validity}. 
If  a correct server $s_i$  ss-delivers a message $m$ from $p$
(reader or writer), then either $p$  ss-broadcasts $m$, or $m$ belongs 
to the initial state of the corresponding link.  
\vspace{-0.2cm}
\item \emph{Order delivery}. 
Any correct server ss-delivers the  messages ss-broadcast by a process $p_x$
(reader or writer) in the order in which they have been ss-broadcast. 
\end{itemize}

Implementations of such a broadcast abstraction are presented in Section 4.2 
of~\cite{D00}, (see  also~\cite{DDPT11,DHSS12}). They rely on bounded 
capacity communication links\footnote{Roughly speaking, 
in a simple implementation, when a message $m$ send operation is invoked 
by a correct process $p_i$ to a correct process $p_j$, $p_i$
repeatedly send the packet $(0,m)$ to $p_j$ until receiving $(cap+1)$ 
packets from $p_j$ (where $cap$ is the maximal  number of packets in 
transit from $p_i$ to $p_j$ and back). Then $p_i$ repeatedly sends 
the packets $(1,m)$ to $p_j$ until receiving $(cap+1)$ packets from $p_j$.
Process $p_j$ sends $(bit,ack)$ only when receiving $(bit,m)$,
and executes ${\sf ss\_deliver}(m)$ when receiving the packet $(1,m)$ 
immediately after receiving the packet $(0,m)$.}.

\subsection{Problem Statement}
\label{sec:pb}
\paragraph{Construction of a read/write register and assumptions}
The problem in which we are interested is the construction of a 
stabilizing server-based  atomic register $\REG$, that can be written 
by the writer $p_w$, and read by the reader $p_r$. 
From an abstraction point of view, the register  provides the writer with
an operation ${\sf write}(v)$, where the input parameter $v$ is the new 
value of the register, and the reader with an  operation ${\sf read}()$,
which returns the value of the register. 

The construction is done incrementally. A regular register is first built.
Then this construction is enriched to obtain an atomic register. 
Both constructions assume that (a) there is a time after which there is no 
more transient failures (instant $\tau_{no\_tr}$),
and (b) the writer invokes at least once the 
${\sf write}()$ operation after $\tau_{no\_tr}$.   
According to  case (b), let $\tau_{1w}> \tau_{no\_tr}$ be the time 
at which the first write invoked after $\tau_{no\_tr}$ terminates.

\paragraph{Concurrent operations, read and write sequences}
Let $W$ and $R$  be the executions of a $\REG.{\sf write}()$ operation by 
the writer and $\REG.{\sf read}()$ operation by the reader, respectively. 
If $W$ and $R$ overlap in time, they are said to be {\it concurrent}. 
If they do not overlap, they are said to be {\it sequential}. 

Let  us observe that, as the writer $p_w$ (resp., reader $p_r$)
is sequential, the set of invocations of the operation ${\sf write}()$  
(resp., ${\sf read}()$) defines a sequence $S_W$ (resp., $S_R$). 

\paragraph{Stabilizing regular register}
A {\it regular} read/write register is  defined by the following 
properties\footnote{These  definitions of a stabilizing regular register, 
and a  stabilizing atomic register, are straightforward extensions of 
the basic definitions given in~\cite{L86-a}.}. 
\begin{itemize}
\vspace{-0.2cm} 
\item \emph{Liveness}. 
Any invocation of $\REG.{\sf write}()$ or $\REG.{\sf read}()$ terminates.
\vspace{-0.2cm} 
\item \emph{Eventual regularity}.
There is a finite time  $\tau_{stab}>\tau_{1w}$
after which each $\REG.{\sf read}()$ $R$ returns a value  $v$  that 
was written by a $\REG.{\sf write}()$ operation $W$ that is 
(a) the last write operation executed before $R$, or 
(b) a write operation concurrent with $R$. 
\end{itemize}
Let us observe that, as there is at least one invocation of 
$\REG.{\sf write}()$ (assumption), and any invocation of 
$\REG.{\sf write}()$ terminates (liveness), $\tau_{1w}$ exists. 
Let us also observe that, before  $\tau_{stab}$, read operations 
can return arbitrary values.  
If a  read/write register is regular, we say that the value returned by 
each of its read operations is regular.

\begin{figure}[th]
\centering{
\vspace{0.1cm}
\scalebox{0.4}{\input{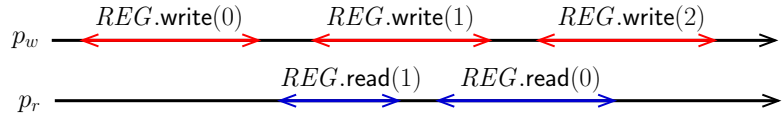}}
\vspace{-0.2cm}
\caption{Regular register: new/old inversion}
\label{fig:new-old-inversion}
}
\end{figure}

The duration $\tau_{stab}-\tau_{no\_tr}$ is the time needed for 
the system to stabilize.  After $\tau_{stab}$, 
no invocation of  $\REG.{\sf read}()$ returns an arbitrary value.  
But, while after  $\tau_{stab}$ regularity prevents $\REG$ from  
returning too ``old'' values,  it still allows  $\REG$ to return values 
in an order different from their writing order, as described 
in Figure~\ref{fig:new-old-inversion}. 
The first read returns the value $1$ (whose write is concurrent with it), 
while the second read returns the value $0$ (which was the last value
written before it starts). 
This phenomenon is known under the name ``new/old inversion''. 

\paragraph{Stabilizing atomic register}
Such a register is a stabilizing regular register that, after some time, 
does not allow new/old inversion. 
It is defined by the following properties. 
\begin{itemize}
\vspace{-0.2cm}
\item \emph{Liveness}.  
Any invocation of $\REG.{\sf write}()$ or $\REG.{\sf read}()$ terminates.
\vspace{-0.2cm} 
\item \emph{Eventual atomicity}. There is a finite time  $\tau_{stab}>\tau_{1w}$
after which any invocation of $\REG.{\sf read}()$ returns a regular value, and 
there are no two invocations of $\REG.{\sf read}()$ that return new/old 
inverted values. 
\end{itemize}
Informally, this means that it is possible to merge sequences $S_W$ and 
$S_R$ to obtain a sequence $S$ where, after time $\tau_{stab}$, each read 
operation  returns the last value written by the closest write operation 
that precedes it. 

\paragraph{Notation and other read/write registers}
The previous registers are called stabilizing regular (or atomic)
single-writer single-reader (SWSR) registers. 
The SWSR atomic register will be used in Section~\ref{sec:single-to-multi}
as a building block to construct
stabilizing  atomic single-writer multi-reader (SWMR) registers, and 
stabilizing  atomic multi-writer multi-reader (MWMR) registers.

\vspace{-0.2cm}
\section{Construction of a Stabilizing SWSR Regular Register}
\label{sec:regular-algorithm}
\vspace{-0.2cm}
This section presents a  stabilizing algorithm that implements a  
single-writer single-reader regular register in the system model introduced 
in Section~\ref{sec:model}.

\subsection{Algorithm}
The algorithms implementing the operations $\REG.{\sf write}()$,
$\REG.{\sf read}()$, and the behavior of the servers $s_i$, is 
described in Figure~\ref{algo:SWSR-regular}. 
The writer and the reader terminate their operations when 
they execute the statement ${\sf return}()$
(line~\ref{ns-byz-06} for the writer, and lines~\ref{ns-byz-13}
or~\ref{ns-byz-15} for the reader).

\paragraph{Local variables and update  messages} 
Each server $s_i$, $1\leq i\leq n$, manages two local variables, which locally
define its internal representation of the constructed regular register $\REG$.

\begin{itemize}
\vspace{-0.2cm}
\item The aim of the variable $last\_val_i$ is to store the
last value written by the writer, as  known by $s_i$. To that end, when
it invokes $\REG.{\sf write}(v)$, the writer ss-broadcasts the message  
{\sc write} $(v)$  to inform the servers of the new value $v$. 
\vspace{-0.2cm}
\item  The aim of the variable $helping\_val_i$  is to contain the
last value ss-broadcast by the writer to each server $s_i$, when 
identifying that the reader requests assistance as write operations are too 
frequent. This variable is reset to $\bot$ at the beginning of every new read.
\end{itemize}

There is no specific local variable managed by  the writer. 
As far as the reader is concerned, it has to manage a single local variable.
\begin{itemize}
\vspace{-0.2cm}
\item
 $\mathit{new\_read_r}$ is a Boolean flag, that, when true, 
demands each server to reset to $\bot$ its  helping variable $helping\_val_i$. 
To this end, the reader ss-broadcasts the message 
{\sc read}$(\mathit{new\_read_r})$, where  $\mathit{new\_read_r}={\tt true}$, 
each time it starts a new read operation. 
\end{itemize}

\begin{figure}[h]
\begin{myalgo}

{\bf operation} $\mathsf{write}$ ($v$) {\bf is}   
                    ~~~~\% issued by the writer $p_w$  \% \\
\line{ns-byz-01} \>  
${\sf ss\_broadcast}$ {\sc write} $(v)$ ${\sf to}$  all servers;\\

\line{ns-byz-02} \> ${\sf wait}$ 
      \big(messages {\sc ack\_write} $(helping\_val)$
                      received from $(n-t)$ different servers\big);\\

\line{ns-byz-03} \> {\bf if} ~\=
$\neg$\big($\exists$ $w\neq \bot$ such that $helping\_val=w$ for $(4t+1)$ 
                                   of the previous messages\big)\\

\line{ns-byz-04} \>\>\>
{\bf then}  
   ${\sf ss\_broadcast}$ {\sc new\_help\_val}$(v)$ ${\sf to}$  all servers\\

\line{ns-byz-05} \> {\bf end if};\\

\line{ns-byz-06} \> ${\sf return}()$.\\~\\

{\bf operation} ${\sf read}$ () {\bf is}   
                    ~~~~\% issued by the reader $p_r$  \% \\

\line{ns-byz-07} \> $\mathit{new\_read_r \leftarrow  {\tt true}}$; \\

\line{ns-byz-08} \> {\bf wh}\={\bf ile}  (${\tt true}$)   {\bf do}\\

\line{ns-byz-09} \>\> 
   ${\sf ss\_broadcast}$ 
{\sc read} $(\mathit{new\_read_r})$ ${\sf to}$    all servers;\\

\line{ns-byz-10} \>\> 
    $\mathit{new\_read_r \leftarrow  {\tt false}}$;  \\

\line{ns-byz-11}\>  \> ${\sf wait}$ 
 \big(messages {\sc ack\_read} $(last\_val,helping\_val)$ \\ 
 \>  \>~~~~~~~~~~~~~~~~~~~~~~~~~~                     
received from $(n-t)$ different servers\big);\\

\line{ns-byz-12} \>  \> {\bf if} \=
  \big($(2t+1)$ of the previous messages
have the same $last\_val$\big)\\
          
\line{ns-byz-13} \>\> \> {\bf then} \=
{\bf let}  $ v$ be this value;  ${\sf return}(v)$

~~~~~ \% the value returned is regular or atomic  \%   \\

\line{ns-byz-14} \> \>\> {\bf else} \> 
    {\bf if} \= \big($(2t+1)$  of the previous messages
have the same $helping\_val\neq \bot$\big)\\

\line{ns-byz-15}\>   \>\>\>\> {\bf then} \=
{\bf let}  $w$  be this value;  ${\sf return}(w)$   
    ~ \% the value returned is atomic \%  \\

\line{ns-byz-16}\> \>  \>\>  {\bf end if}\\

\line{ns-byz-17} \>\>  {\bf end if}\\

\line{ns-byz-18} \> {\bf end while}.\\

-----------------------------------------------------------------------------------------------------------\\

{\bf when} {\sc write} $(v)$ {\bf is} 
                 ${\sf ss\_delivered}$  {\bf from} $p_w$ {\bf do}\\

\line{ns-byz-19} \> $last\_val_i\leftarrow v$; \\

\line{ns-byz-20} \> 
${\sf send}$  
{\sc ack\_write} $(helping\_val_i)$ ${\sf to}$ $p_w$.
\\~\\

{\bf when} {\sc new\_help\_val} $(v)$ 
      {\bf is}   ${\sf ss\_delivered}$  {\bf from} $p_w$ {\bf do}\\

\line{ns-byz-21} \>   $helping\_val_i\leftarrow v$.\\~\\

{\bf when}  {\sc read} $(\mathit{new\_read})$
          {\bf is} ${\sf ss\_delivered}$  {\bf from} $p_r$ {\bf do}\\

\line{ns-byz-22} \> {\bf if} $(\mathit{new\_read})$
     {\bf then} $helping\_val_i \leftarrow \bot$ 
     {\bf end if};\\

\line{ns-byz-23} \> 
  ${\sf send}$  {\sc ack\_read} 
     $( last\_val_i,helping\_val_i)$ ${\sf to}$ $p_r$.

\end{myalgo}
\caption{Byzantine-tolerant stabilizing  SWSR regular register}
\label{algo:SWSR-regular}
\end{figure}

\paragraph{Algorithm implementing  $\REG.{\sf write}()$}
As already said, when the writer invokes $\REG.{\sf write}(v)$, it first
ss-broadcasts the message {\sc write}$(v)$ (line~\ref{ns-byz-01}), 
and  waits until it received an acknowledgment message  
{\sc ack \_write}$(helping\_val)$ from $(n-t)$ servers, (i.e., from at least 
$(n-2t)$ correct servers)  (line~\ref{ns-byz-02}).

When a server $s_i$ ss-delivers  the message {\sc write}$(v)$, it updates
$last\_val_i$ (line~\ref{ns-byz-19}), and sends by return (line~\ref{ns-byz-20})
the acknowledgment {\sc ack \_write}$(helping\_val_i)$ to give the writer
information on the state of the reader (namely, $helping\_val_i=\bot$ 
means that the reader started a new read operation, and accordingly
 $helping\_val_i$ needs to be refreshed). 

When the writer stops waiting, it checks if it has received the same value
$helping\_val\neq \bot$ from at least $(4t+1)$ different servers
(line~\ref{ns-byz-03}).
If this predicate is false, the local variables $helping\_val_i$ 
of the servers $s_i$ needs to be refreshed. To this end, the writer
 ss-broadcasts the message {\sc new\_help\_val}$(v)$ to inform them that, 
from now on, they must consider $v$ as the new helping value 
(lines~\ref{ns-byz-04}  and~\ref{ns-byz-21}).

\paragraph{Algorithm implementing  $\REG.{\sf read}()$}
When the reader invokes  $\REG.{\sf read}()$, it sets $new\_read_r$ to 
$\mathit{\tt true}$ (line~\ref{ns-byz-07}) and enters a while loop
(lines~\ref{ns-byz-08} and~\ref{ns-byz-18}), that it will exit at 
line~\ref{ns-byz-13} or~\ref{ns-byz-15}. 
Once in the loop body, the reader  starts a new inquiry by 
ss-broadcasting the message {\sc read}$(new\_read_r)$ to the servers.  
If $\mathit{new\_read_r={\tt true}}$, the message 
is related to a new read operation (line~\ref{ns-byz-07}); 
if $\mathit{new\_read_r={\tt false}}$,
it is from the same read operation as before (line~\ref{ns-byz-10}). 
Then, the reader   waits until it has  received  an
 acknowledgment message {\sc ack \_read}$(last\_val,helping\_val)$ 
from $(n-t)$  servers (line~\ref{ns-byz-11}).

When a server $s_i$  receives  the message {\sc read}$(new\_read_r)$, 
it resets $helping\_val_i$ to $\bot$ if this message 
indicates a new read operation started (line~\ref{ns-byz-22}). 
In all cases (i.e., whatever the value of $new\_read_r$),  
it sends by return its current local state in the message 
{\sc ack \_read}$(last\_val_i,helping\_val_i)$ (line~\ref{ns-byz-23}). 

When the reader stops waiting, it returns the value $v$ if 
the field  $last\_val$ of $(2t+1)$ messages {\sc ack \_read}$()$ 
is equal to $v$ (lines~\ref{ns-byz-12}-\ref{ns-byz-13}). 
Otherwise it  returns the value $w$ if  
the field  $helping\_val$ of $(2t+1)$ messages {\sc ack \_read}$()$ 
is equal to $w\neq\bot$  (lines~\ref{ns-byz-14}-\ref{ns-byz-15}). If none of 
these predicates is satisfied, the reader re-enters the loop body.

\paragraph{Remark on the reception order of the  messages 
{\sc ack\_write}$()$ and {\sc ack\_read}$()$}
It is important to notice that, thanks to the properties of the ss-broadcast 
abstraction, and the fact that the links are FIFO, we have the following. 
When the writer invokes ${\sf ss\_broadcast}()$, and later waits for  
associated acknowledgments  {\sc ack\_write}$()$ from $(n-t)$ servers
(lines~\ref{ns-byz-01}-\ref{ns-byz-02}), the sequence of 
acknowledgments received from each correct server matches the sequence of 
${\sf ss\_broadcast}()$ it issued (the same holds for the reader and  the 
acknowledgments {\sc ack\_read}$()$, lines~\ref{ns-byz-09}-\ref{ns-byz-11}).  
This means that ${\sf ss\_broadcast}()$ and the associated  acknowledgments 
do not need to carry sequence numbers.

\subsection{Proof of the construction}
\label{sec:basic-algo-proof}
All the poofs assume $n\geq 8t+1$.
\begin{lemma}
\label{write-termination}
Any invocation of $\mathsf{write}~()$ terminates.
\end{lemma}
\vspace{-0.2cm}
\begin{proofL}
Due to the ss-broadcast termination property, the writer cannot block forever
when it invokes ${\sf ss\_broadcast}()$ at line~\ref{ns-byz-01} or 
line~\ref{ns-byz-04}. As far the wait statement of line~\ref{ns-byz-02} 
is concerned, we have the following: due to the ss-broadcast eventual delivery 
property, eventually at least $(n-t)$ non-Byzantine servers ss-deliver 
the message  {\sc write}$()$ ss-broadcast by the writer,  
and then they will eventually answer by  returning the acknowledgment 
message {\sc ack\_write}$()$, which concludes the proof of the lemma.  
\renewcommand{\toto}{write-termination}
\end{proofL}

\begin{lemma}
\label{read-termination}
Any invocation of $\mathsf{read}~()$ terminates. 
\end{lemma}

\begin{proofL}
Using the same reasoning as in Lemma~\ref{write-termination}, 
it follows that the reader cannot block forever in the wait statement of 
line~\ref{ns-byz-11}. So, the proof consists in showing that the predicate of 
line~\ref{ns-byz-12}, or the one of~\ref{ns-byz-14}, becomes eventually true. 
The rest of the proof is by contradiction. Let $R$ be the first invocation 
of $\mathsf{read}~()$ that does not terminate and let us consider an 
execution of the loop body after time $\tau_{stab}$. 

{\sl Claim} C. At the time at which a write that started after $\tau_{no\_tr}$ 
terminates, there are (a) at least $(n-2t)$ correct servers $s_i$ such that 
$last\_val_i=v$, and (b) at least $(3t+1)$ correct servers $s_j$ such that 
$helping\_val_j=w\neq\bot$.\\
{\sl Proof of the claim}.
Let us consider a write started after $\tau_{no\_tr}$ and let $\tau_{w}$ 
be the time at which such write terminates.
Considering that after $\tau_{no\_tr}$ there are no more transient failures
and due to the synchronized delivery property of the ss-broadcast we have that 
at time $\tau_{w}$ there are at least $(n-2t)$ correct servers $s_i$ such that 
$last\_val_i=v$.
Moreover, if the predicate of line~\ref{ns-byz-03} is true, it follows from
(a) the synchronized delivery property of the ss-broadcast of the message
{\sc new\_help\_val}$()$ (line~\ref{ns-byz-04}), and (b) the fact that 
$n-2t>3t+1$, that at least $(3t+1)$ correct servers $s_j$ are such that
$helping\_val_j=w\neq\bot$.
If predicate of line~\ref{ns-byz-03} is false, there are $(4t+1)$
servers that sent {\sc ack\_write}$(w)$ where $w\neq \bot$
(line~\ref{ns-byz-20}),
from which we conclude that at least $(3t+1)$ of them are correct and are
such that $helping\_val_j=w\neq\bot$. {\sl End of the proof of the claim} C.

Let us consider the last  write that terminated before $R$ 
started,  and let us assume it wrote $x$.  Due to part (a) of Claim C,
just after this write terminated, at least $(n-2t)$ correct servers $s_i$ 
are such that $last\_val_i=x$.  If no write is concurrent with $R$, as $R$
receives messages {\sc ack\_read}$(last\_val,-)$ from $(n-t)$ 
servers at line~\ref{ns-byz-11} (i.e., from at least $(n-2t)$ correct servers), 
it follows from the fact that the intersection of any two sets 
$Q1$ and $Q2$ of $(n-2t)$  correct servers (the set $Q1$ of correct servers 
$s_i$ such that $last\_val_i=x$, and the set $Q2$ of correct servers from 
which $R$ receives  {\sc ack\_read}~$(last\_val,-)$) contains at least 
$(2t+1)$  correct servers, that $R$ terminates at 
lines~\ref{ns-byz-12}-\ref{ns-byz-13}.

Let us now assume that there is exactly one write that is 
concurrent with $R$, and  let $y$ be the value it writes.
Due to the synchronized delivery property of ss-broadcast, $R$ first resets 
to $\bot$ the variables $helping\_val_i$ of at least $n-2t\geq 6t+1$ 
correct servers $s_i$ (lines~\ref{ns-byz-07},~\ref{ns-byz-09}, 
and~\ref{ns-byz-22}), and then receives (line~\ref{ns-byz-11}) messages 
{\sc ack\_read}$(last\_val,-)$ from at least $n-2t\geq 6t+1$ correct servers.
We show that at least $(2t+1)$ of these messages carry either $x$ or $y$, 
from which $R$ terminates at lines~\ref{ns-byz-12}-\ref{ns-byz-13}.  
Due to part (a) of Claim C, there were at least $n-2t\geq 6t+1$ correct 
servers $s_i$ such that $last\_val_i=x$  when the write of $x$ finished.
Let $Q$ be this set of servers.
$R$ receives messages {\sc ack\_read}$(last\_val,-)$  from at least 
$(4t+1)$ servers in $Q$.
Due to the operation ${\sf write}(y)$ (concurrent with $R$), variables
$last\_val_i$ of some of these servers may have been updated to the value $y$.
Hence, some of the previous $(4t+1)$ messages {\sc ack\_read}$(last\_val,-)$
received by $R$ carry $x$, while others carry $y$. Hence, at least $(2t+1)$
of them carry either $x$ or $y$, and $R$ terminates
at lines~\ref{ns-byz-12}-\ref{ns-byz-13}.

Let us finally consider the case where there are more than one write 
concurrent with $R$. When $R$ terminates its  invocation of
${\sf ss\_broadcast}$ {\sc read}$({\tt true})$  
(there is only one such invocation per read operation, line~\ref{ns-byz-09}), 
the local variables $helping\_val_i$ of $(n-2t)$ correct servers are equal 
to $\bot$. Let $Q'$ be this set of servers. 
(The proof of this statement is the same as the proof appearing in the
first part of  claim C.) 
Hence, when this ss-broadcast terminated,  the messages 
{\sc ack\_read}$(-,helping\_val_i)$  sent by each  server $s_i\in Q'$
(line~\ref{ns-byz-23}),  is such that $helping\_val_i=\bot$. 
Let us consider the first write (e.g., ${\sf write}(z)$)
that occurs after the servers $s_i\in Q'$
have set $helping\_val_i$ to $\bot$. This write receives $(n-t)$
messages {\sc ack\_write} $(helping\_val)$, and at least
$(4t+1)$ of them are from servers in $Q'$ and carry  $helping\_val=\bot$.  
Hence the predicate of line~\ref{ns-byz-03} is satisfied, and the writer issues
${\sf ss\_broadcast}$~{\sc new\_help\_val}$(z)$. 
If later (i.e.,  after the invocation of  ${\sf write}(z)$ terminated),
there are other invocations of  ${\sf write}()$ concurrent with $R$, 
none of them will execute line~\ref{ns-byz-04}. This is due to the fact that 
$R$ does not reset  the variables $helping\_val_i$ to $\bot$, 
and the $(n-t)$ messages {\sc ack\_write}~$(helping\_val)$ sent by the servers
at line~\ref{ns-byz-20} are such that at most $t$ are from Byzantine servers, 
and at least $(4t+1)$ carry $z$, from which 
follows that there is a finite time $\tau_R$ after which the variables 
$helping\_val_i$ of the correct servers are no longer modified. Let us 
finally consider the first invocation of ${\sf ss\_broadcast}~(new\_read_r)$
issued by $R$ after  $\tau_R$,  such that  $new\_read_r={\tt false}$.
It follows from the previous discussion that, among the 
$(n-t)$ messages {\sc ack\_read}$(-,helping\_val)$ received by $R$,
at most $t$ (the ones from Byzantine servers) carry arbitrary values, 
and at least $(n-t)-3t \geq 4t+1$ carry the value $z$.  When this occurs, 
$R$ terminates at  lines~\ref{ns-byz-14}-\ref{ns-byz-15}.
\renewcommand{\toto}{read-termination}
\end{proofL}

\begin{lemma}
\label{label:eventual-regularity}
Let $t<n/8$.
There is a finite time  $\tau_{stab}>\tau_{1w}$
after which each read invocation $R$ returns a value  $v$  that 
was written by a write operation $W$, which is 
(a) the last write operation executed before $R$, or 
(b) a write operation concurrent with $R$. 
\end{lemma}

\begin{proofL}
Let us assume that a read operation $R$
 returns $z$,  a value  different from
the value $v$ of the  last completed write prior to $R$,  
and from any  value $u$ of a concurrent write. 
Let us consider the first write concurrent with $R$. For $R$ to return
$z$, the reader must receive $(2t+1)$ messages
{\sc ack\_read}$(z,-)$ or $(2t+1)$ messages {\sc ack\_read}$(-,z)$.
However, immediately following the termination of the write of $v$ there 
were $(n-2t)$ correct servers $s_i$ 
with $last\_val_i=v$. Thus,  following the termination of the write of $v$,
and until the termination of the next write of some value $u$,
the reader cannot receive 
$(2t+1)$ values for a value  $z$ different from $v$ and $u$. 
The above argument holds for the second concurrent write, where we start 
with $(n-2t)$ values of $u$, and so on and so forth.
\renewcommand{\toto}{label:eventual-regularity}
\end{proofL}
 
\begin{theorem}
\label{theorem:regular-register}
Let $t<n/8$. The algorithm described in Figure~{\em\ref{algo:SWSR-regular}}
implements a stabilizing regular register in the presence of at 
most $t$ Byzantine servers.  
{\em (The proof follows from Lemmas~\ref{write-termination},
\ref{read-termination}, and~\ref{label:eventual-regularity}.)}
\end{theorem}

\subsection{The case of synchronous links}
\label{sec:synchronous-SWSR}
Let us consider a communication model where the links are synchronous. 
{\it Synchronous} means here that  each link, connecting the
reader or the writer and a correct server, is timely
i.e.,  there is an upper bound on message transfer delays  and this 
bound is known by the processes. When considering the construction of 
an SWSR regular register, this allows the reader or the writer to know 
how long it has to wait for a round trip delay with respect to the  
correct servers, and consequently use a timer with an appropriate timeout value.

It appears that the previous algorithm  can be adapted, 
with very a simple modification, to this synchronous communication model to 
build a stabilizing SWSR regular register. Due to page limitation, this 
algorithm is described and proved correct in Appendix~\ref{SWSRSS}. 
The important result is the following theorem, which states that, in 
such a synchrony setting, up to $t<n/3$ servers can commit Byzantine failures.

\begin{theorem}
\label{theorem:regular-register-ss-2}
Let $t<n/3$. The algorithm described in 
Figure~{\em\ref{algo:SWSRSS-regular-ss}}
implements a stabilizing regular register in the presence of at 
most $t$ Byzantine servers.  {\em (Proof in Appendix~\ref{SWSRSS}.)}
\end{theorem}

\section{Construction of a Stabilizing SWSR  Practically Atomic Register}
\label{sec:practically-atomic}

\paragraph{Practically stabilizing SWSR atomic register}
A  stabilizing SWSR  {\it practically}
atomic  register is  a stabilizing SWSR regular register with no 
new/old inversions as long
as the number of writes between two successive reads (that are not executed   
concurrently with any write) is less than a given constant called 
{\it  system-life-span} (e.g., $2^{64}$)~\cite{AADDPT11}.

This section presents a practically stabilizing SWSR atomic register that 
stabilizes  after a read that (a) is not concurrent with a write, and 
(b) follows the first write that follows the last transient failure.
Its operations are denoted ${\sf prac\_at\_write}$() 
and ${\sf prac\_at\_read}$().

\paragraph{Algorithm}
The stabilizing SWSR  {\it practically}
atomic  register algorithm is described in Figure~\ref{algo:SWSR-atomic}.
It is an extension of the algorithm implementing a stabilizing regular 
register presented   Figure~\ref{algo:SWSR-regular}. 
The lines with the same number $xy$ are exactly the same in both algorithms.  
A line numbered N$x$ is a new line, while a line numbered $xy$M$z$ corresponds 
to a modification of the line $xy$  of Figure~\ref{algo:SWSR-regular}.

\paragraph{Underlying principle}
To obtain an  algorithm implementing such a register, 
the main idea is to count the invocations of  ${\sf prac\_at\_write}$() 
so that no new/old inversion can occur if the reader traces the sequence 
number attached to each written value, and exchange an older value with a 
newer  that is already known. This is the role of the write sequence number 
denoted $\underline{wsn}$. Hence, the data value $v$
appearing in Figure~\ref{algo:SWSR-regular} in 
now replaced by  the pair  $(\underline{wsn},v)$ in  
Figure~\ref{algo:SWSR-atomic}.
Therefore, $last\_val_i$ contains now such a pair, and 
 $helping\_val_i$ contains now either such a pair, or the default value $\bot$.

Special care must be taken to bound   $\underline{wsn}$
so that there is no ambiguity on its current value. 
Hence, a relation $\succeq$  on sequence numbers has to be defined, 
such that it always reflects the write order of the values 
they are associated with.  
This relation is defined as follows: 
given two integer $x$ and $y$ (e.g., in range $[0, 2^{128}+1]$), 
$x \geq_{cd} y$  iff the clockwise distance (hence the subscript $cd$)
from $y$ to  $x$ is smaller than their anti-clock distance; moreover, 
$x >_{cd} y$ if $x \geq_{cd} y$ and $x\neq y$. 
Such precedence relation is used at lines~N6 and \ref{ns-byz-13}M2 to
 compare the highest previously received sequence number $pwsn$ with 
the current one and to 
update it (lines N6, \ref{ns-byz-13}M2, and \ref{ns-byz-15}M).
As transient failures may corrupt  counter values, those
must be automatically corrected. This is done as follows.
After the first  read,  which follows a write invocation 
and does not overlap a write, it holds that the local pair $(pwsn,pv)$ stored by
the reader reflects the last read correct value. Thus, the bookkeeping of 
$pwsn$,  $pv$, and the values of $\underline{wsn}$ and $v$, which are read, 
reflects the right value ordering which  allow their correct reordering, 
thereby  providing the writer and the reader with  an atomic register. 

The aim of the lines  N2-N7 is to do a sanity check for the 
the local pair $(pwsn,pv)$ managed by the reader. 
To that end, the reader ss-broadcasts the message {\sc read} $({\tt false})$, 
and wait for $(n-t)$ associated acknowledgments 
{\sc ack\_read} $(-,helping\_val)$ (lines~N2-N3). 
If $(2t+1)$ of these messages carry the same pair 
$helping\_val=(\underline{wsn},v)$, and  $\underline{wsn}$ is smaller than 
$pvsn$, then the reader adopts this pair as current value of $(pvsn,pv)$. 
This is because, if  $(2t+1)$ of these messages carry the same pair, they 
reflect the last value written, and therefore  carry the
correct $wsn$. Hence, the ``if'' statement in line~N6, whose aim is to 
refresh the pair $(pvsn,pv)$. 
This preliminary sanity check, which relies on values provided by the servers, 
helps the rest of the read algorithm
(lines~\ref{ns-byz-07}-\ref{ns-byz-18} which are nearly the same as the ones 
of Figure~\ref{algo:SWSR-regular}) prevent new/old inversions from occurring.

\paragraph{Remark} 
Due to page limitation, the proof of the previous construction is given 
in Appendix~\ref{annex-proof-practically-atomic-SWSR}. Let us notice that
 the  ``synchronous link'' algorithm designed for $n \geq 3t+1$ processes, 
has a similar extension, which builds an  SWSR atomic register version. 


\begin{figure}[h!]
\begin{myalgo}
{\bf operation} $\mathsf{prac\_at\_write}$ ($v$) {\bf is}   ~~~~\% issued by the writer $p_w$  \% \\

(N1) \>  ~~$\underline{wsn} \leftarrow (\underline{wsn}+1) \mbox{ mod }(2^{64}+1)$;\\

(\ref{ns-byz-01}M) \>   ~ ~${\sf ss\_broadcast}$ {\sc write} $(\underline{wsn},v)$ ${\sf to}$  all servers;\\

(\ref{ns-byz-02}) \>  ~~${\sf wait}$ \big(messages {\sc ack\_write} $(helping\_val)$ received from $(n-t)$ different servers\big);\\

(\ref{ns-byz-03}) \>   ~~{\bf if} ~\= $\neg$\big($\exists$ $w\neq \bot$ such that $helping\_val=w$ for $(4t+1)$ of the previous messages\big)\\

(\ref{ns-byz-04}M) \>\>\> ~~{\bf then}  ${\sf ss\_broadcast}$ {\sc new\_help\_val}$(\underline{wsn},v)$ ${\sf to}$  all servers\\ 

(\ref{ns-byz-05}) \>  ~{\bf end if}\\

(\ref{ns-byz-06}) \>  ~${\sf return}()$.\\~\\

{\bf operation} ${\sf prac\_at\_read}$ () {\bf is} ~~~~\% issued by the reader $r_i$ where $1\leq i\leq r$  \% \\

(N2) \>  ~${\sf ss\_broadcast}$ {\sc read} $({\tt false})$ ${\sf to}$    all servers;\\

(N3) \> ~ ${\sf wait}$ \big(messages {\sc ack\_read} $(last\_val,helping\_val)$ 
received from $(n-t)$ different servers\big);\\
 
(N4) \> ~{\bf if} \= \big($(2t+1)$  of the previous messages have the same $helping\_val\neq \bot$\big)\\

(N5) \> \>  ~{\bf then} \= {\bf let}  $(\underline{wsn}, v)$ be this value;\\

(N6)  ~\>\>\> {\bf if} $(pwsn >_{cd} \underline{wsn})$ {\bf then} $pwsn \leftarrow \underline{wsn}$; $pv\leftarrow v$ {\bf end if} ~\% sanity check for $pwsn$ and $pv$  \%\\ 

(N7) \> ~{\bf end if};\\

(\ref{ns-byz-07}) ~\> $\mathit{new\_read_r \leftarrow  {\tt true}}$; \\

(\ref{ns-byz-08}) \> ~{\bf wh}\={\bf ile}  (${\tt true}$)   {\bf do}\\

(\ref{ns-byz-09}) \>\> ~${\sf ss\_broadcast}$ {\sc read} $(\mathit{new\_read_r})$ ${\sf to}$    all servers;\\

(\ref{ns-byz-10}) \>\> ~$\mathit{new\_read_r \leftarrow  {\tt false}}$;  \\

(\ref{ns-byz-11})\>  \> ~ ${\sf wait}$ \big(messages {\sc ack\_read} $(last\_val,helping\_val)$ \\ 
 \>  \>~~~~~~~~~~~~~~~~~~~~~~~~~~ received from $(n-t)$ different servers\big);\\

(\ref{ns-byz-12})\>  \> ~{\bf if} \= \big($(2t+1)$ \= of the \= previous \=  messages have the same $last\_val$\big)\\
          
(\ref{ns-byz-13}M1)  \>\> \> ~{\bf then} \= {\bf let}  $(\underline{wsn}, v)$ be this value;\\

(\ref{ns-byz-13}M2) \>\>\>\> {\bf if} $(\underline{wsn} >_{cd} pwsn)$ \= {\bf then} \= $pwsn \leftarrow \underline{wsn}$; $pv \leftarrow v$;  ${\sf return}(v)$ \\

(\ref{ns-byz-13}M3) \>\>\>\> \> {\bf else}\> ~${\sf return}(pv)$ ~ \% prevention of new/old inversion \%\\

(\ref{ns-byz-13}M4) \>  \> \>  \>  ~{\bf end if} \\

(\ref{ns-byz-14}) \> \>\> {\bf else} \>  {\bf if} \= \big($(2t+1)$  of the previous messages have the same $helping\_val\neq \bot$\big)\\

(\ref{ns-byz-15}M) \>   \>\>\>\> ~{\bf then} \= {\bf let}  $(\underline{wsn},w)$  be this value; $pwsn \leftarrow \underline{wsn}$; $pv \leftarrow w$;  ${\sf return}(w)$
    ~ \% already atomic \%  \\

(\ref{ns-byz-16})\> \>  \>\>  ~{\bf end if}\\

(\ref{ns-byz-17}) \>\>  ~{\bf end if}\\

(\ref{ns-byz-18}) \> ~{\bf end while}.\\

--------------------------------------------------------------------------------------------------------------------------------------\\

{\bf when} {\sc write} $(sn, v)$ {\bf is} ${\sf ss\_delivered}$  {\bf from} $p_w$ {\bf do}
            \% $v$ is now a pair $(seq.~nb, value)$ \% \\

(\ref{ns-byz-19})  \> ~$last\_val_i\leftarrow v$; \\

(\ref{ns-byz-20}) \> ${\sf send}$ {\sc ack\_write} ~$(helping\_val_i)$ ${\sf to}$ $p_w$.\\~\\

{\bf when} {\sc new\_help\_val} $(v)$ {\bf is}   ${\sf ss\_delivered}$  {\bf from} $p_w$ {\bf do}
          \% $v$ is now a pair $(seq.~nb, value)$ \% \\

(\ref{ns-byz-21}) \>   ~$helping\_val_i\leftarrow v$.\\~\\

{\bf when}  {\sc read} $(\mathit{new\_read})$ {\bf is} ${\sf ss\_delivered}$  {\bf from} $p_r$ {\bf do}\\

(\ref{ns-byz-22}) \> ~{\bf if} $(\mathit{new\_read})$ {\bf then} $helping\_val_i \leftarrow \bot$ {\bf end if};\\

(\ref{ns-byz-23}) \> ${\sf send}$  {\sc ack\_read} $( last\_val_i,helping\_val_i)$ ${\sf to}$ $p_r$.

\end{myalgo}
\caption{Byzantine-tolerant practically stabilizing 
         SWSR atomic register}
\label{algo:SWSR-atomic}
\end{figure}

\vspace{0.2cm}
\section{Construction of  Stabilizing SWMR and MWMR Atomic Registers}
\label{sec:single-to-multi}

\vspace{0.1cm}
\subsection{Construction of a  Stabilizing SWMR  Atomic Register}
\label{sec:swmr-algorithm}
The technique to obtain a  SWMR  atomic register from  SWSR  atomic registers
is a classical one~\cite{L86-b,R13}.
The writer interacts with each reader, writing the same value to all readers, 
the servers maintaining variables for each reader.
Since the result is atomic register for each reader, and any write is executed 
to all, then the result is a single-writer multi-reader register.  
Let  $\mathsf{swmr\_write}$() and  $\mathsf{swmr\_read}$() denote the 
operations associated with such a SWMR atomic register.

\subsection{Construction of a Stabilizing MWMR Atomic Register}
\label{sec:mwmr-algorithm}

This section presents a stabilizing algorithm that implements a
multi-writer multi-reader atomic register in the system model introduced
in Section~\ref{sec:model}.
%

\paragraph{Underlying SWMR atomic registers}
It is assumed that each process is both a reader and a writer. 
Hence, in the following we use the term ``process''. 
Let $m$ be the number of processes. 
A process is denoted $p_i$, $1\leq i\leq m$.
The construction uses one stabilizing SWMR register per process. 
Let $\REG[i]$ be the SWMR register associated with $p_i$, 
which means that any process can read it but only $p_i$ can write it.

To write  $\REG[i]$, $p_i$ invokes 
$\REG[i].\mathsf{swmr\_write}\big(v, epoch, seq\big)$,  
where $epoch$ is a bounded label (see below), and 
$seq$ is a sequence number bounded by some large constant $2^{64}$.
Any process $p_j$ reads $\REG[i]$ by invoking $\REG[i].\mathsf{swmr\_read}()$. 
Such an invocation returns a triple $(v, epoch, seq\big)$, where $v$ is a
 data value, whose  associated timestamp is the pair  $(epoch,seq)$.

\begin{figure}[h]

\begin{myalgo}

{\bf operation} $\mathsf{mwmr\_write}$ ($v$) {\bf is}

~~~~\% issued by process  $p_i$ \% \ \\

\line{MWMR-byz-01}\>     
   {\bf  for} $j\in\{1,...,m\}$ 
   {\bf do}  $reg_i[j]\leftarrow \REG[j].\mathsf{swmr\_read}()$  {\bf end for};
    \% obtains $m$ triples  $(val, epoch, seq)$ \% \\ 

\line{MWMR-byz-02} \> {\bf if}
     $\big((\not\exists ~ {\sf max\_epoch}(reg_i[1..m]))$
          $\vee$ $(\exists j:[(reg_i[j].epoch = {\sf max\_epoch}(reg_i[1..m])) 
              \wedge (reg_i[j].seq \geq2^{64})])\big)$\\

\line{MWMR-byz-03}\> \>{\bf then} \= 
 $reg_i[i] \leftarrow \big(v,{\sf next\_epoch}(reg_i[1..m]),0\big)$\\

\line{MWMR-byz-04} \> {\bf end if};\\

\line{MWMR-byz-05}\> let $M$ be the set of indexes $j$ 
                 such that ${\sf max\_epoch}(reg_i[1..m])=reg_i[j].epoch$;\\

\line{MWMR-byz-06}
             \> $seq_{max} \leftarrow {\sf max}(reg_i[j].seq, j \in M)$;\\

\line{MWMR-byz-07}
 \> $\REG[i].{\sf swmr\_write} 
         \big(v,{\sf max\_epoch}(reg_i[1..m]),seq_{max}+1\big)$;\\

\line{MWMR-byz-08} ${\sf return}()$.\\~\\



{\bf operation} ${\sf mwmr\_read}$ () {\bf is}

~~~~\% issued by  process $p_i$ \% \\

\line{MWMR-byz-09}\>     
{\bf  for} $j\in\{1,...,m\}$ 
  {\bf do} $reg_i[j] \leftarrow \REG[j].\mathsf{swmr\_read}()$ {\bf end for};  
    \% obtains $m$  triples $(val, epoch, seq)$ \% \\

\line{MWMR-byz-10} \> {\bf if}
     $\big((\not\exists ~{\sf max\_epoch}(reg_i[1..m]))$ 
                          $\vee$ $(\exists j: [(reg_i[j].epoch = 
     {\sf max\_epoch}(reg_i[1..m])) \wedge (reg_i[j].seq \geq2^{64})])\big)$\\

\line{MWMR-byz-11} \> \>{\bf then} \=
$reg_i[i] \leftarrow (reg_i[i].v,{\sf next\_epoch}(reg_i[1..m]),0\big)$; $\REG[i].{\sf swmr\_write} 
         \big(reg_i[i].v, reg_i[i].epoch, 0\big)$\\

\line{MWMR-byz-12} \> {\bf end if};\\ 

\line{MWMR-byz-13} \> let $M$ be the set of indexes $j$ 
                such that ${\sf max\_epoch}(reg_i[1..m])=reg_i[j].epoch$;\\

\line{MWMR-byz-14} \>$seq_{max} \leftarrow {\sf max}(reg_i[j].seq, j \in M)$;\\

\line{MWMR-byz-15} \> 
let $min \in M$ be the minimal index such that $reg_i[min].seq=seq_{max}$;\\

\line{MWMR-byz-16} \>
 ${\sf return}(reg_i[min].v)$.
\end{myalgo}
\caption{Byzantine-tolerant stabilizing MWMR atomic register 
         from  SWMR  registers}
\label{algo:MWMR-atomic}
\end{figure}
\paragraph{The notion of an epoch} 
This notion was introduced in~\cite{AADDPT11} where a bounded labeling 
scheme is proposed with uninitialized values.  
Let $k > 1$ be an integer, and let $K = k^2+1$.
We consider the set $X=\{1,2,..,K\}$ and let ${\cal L}$
(the set of epochs) be the set of all ordered pairs $(s,A)$
where $s \in X$   and $A \subseteq X$ has size $k$.
 
The comparison operator $\succ$ among two epochs is
defined as follows:
\vspace{-0.3cm}
$$
(s_i,A_i) \succ (s_j,A_j) 
     \stackrel{\mathit{def}}{=} (s_j \in A_i) \wedge (s_i \not \in A_j).
$$
\vspace{-0.1cm}
Note that this operator is antisymmetric by definition,
yet may not be defined for every pair $(s_i,A_i)$ and $(s_j,A_j)$
in ${\cal L}$ (e.g., $s_j \in A_i$ and $s_i \in A_j$).

Given a subset $S$ of epochs of ${\cal L}$, a function is defined 
in~\cite{AADDPT11} which compute
a new epoch which is greater (with respect to $\succ$)
than every label in $S$.
This function, called ${\sf next\_epoch}()$, is as follows.
Given a subset of $k$ epochs $(s_1,A_1), (s_2,A_2), \ldots, (s_k,A_k)$,
 ${\sf next\_epoch}\big((s_1,A_1), (s_2,A_2), \ldots, (s_k,A_k)\big)$ 
is the epoch $(s,A)$ that satisfies:
\begin{itemize}
\vspace{-0.2cm}
\item[--]
$s$ is an element of $X$ that is not in the union
$A_1 \cup A_2 \cup \ldots \cup A_k$ (as the size of each $A_s$ is $k$,
the size of the union is  at most $k^2$,
and since $X$ is of size $k^2+1$ such an $s$ always exists).
\vspace{-0.2cm}
\item[--]
$A$ is a subset of size $k$ of $X$ containing all values 
$(s_1, s_2,\ldots, s_k)$ (if they are not pairwise distinct,
add arbitrary elements of $X$ to get a set of size exactly $k$).
\end{itemize}
\noindent
The relation $\succ$ is  extended  to $\succeq$ as follows:
\vspace{-0.3cm}
$$
(s_i,A_i) \succeq (s_j,A_j) \stackrel{\mathit{def}}{=}
   ((s_i,A_i) \succ (s_j,A_j)) \vee ((s_i=s_j) \wedge (A_i = A_j)).
$$

The predicate ${\sf max\_epoch}()$ applied to a set of epochs returns true if 
there is an epoch in the set such that is equal to or greater (in the sense
 of the relation $\succeq$) than any other epoch in the set.

\paragraph{Algorithm implementing  $\mathsf{mwmr\_write}()$}
When a process $p_i$ invokes  $\mathsf{mwmr\_write}$ ($v$), it 
first checks if it has to start a new epoch 
(lines~\ref{MWMR-byz-01}-\ref{MWMR-byz-04}),
 in which it 
first reads all the underlying SWMR registers $\REG[1..m]$,
and saves their  values in its local array $reg_i[1..m]$ 
(line~\ref{MWMR-byz-01}). This constitutes its view of the global state. 
Hence, for any $j\in\{1,...,m\}$, $reg_i[j]$ contains a triple $(v,epoch,seq)$, 
namely,  $reg_i[j].v$ is the data value of  $\REG_j$, 
$reg_i[j].epoch$ is the epoch of the timestamp of $v$, and 
$reg_i[j].seq$ is  the sequence number of the timestamp of $v$.
 
Then, if there is no greatest epoch in  $reg_i[1..m]$, or there is one
($reg_i[j].epoch$),  but the associated sequence number ($reg_i[j].seq$)
is equal to or greater than the  bound $2^{64}$, $p_i$ must start the next 
epoch ($ne={\sf next\_epoch}(reg_i[1..m]$) with  starts with the sequence 
number $0$, and informs the other processes. To this end $p_i$  writes 
the value $v$ and its timestamp $(ne,0)$ in $\REG_i[i]$.  

Then $p_i$ writes
the value $v$  with its epoch and sequence number (line~\ref{MWMR-byz-07}). 
The pair (epoch, sequence number) is computed at 
lines~\ref{MWMR-byz-05}-\ref{MWMR-byz-07} so that it is greater than 
all the previous pairs known by $p_i$.

\paragraph{Algorithm implementing  $\mathsf{mwmr\_read}()$}
The algorithm implementing the operation  $\mathsf{mwmr\_read}$ () is nearly 
the same as the one  implementing the operation  $\mathsf{mwmr\_write}$ (). 
The lines~\ref{MWMR-byz-09}-\ref{MWMR-byz-12} are the same as the 
lines~\ref{MWMR-byz-01}-\ref{MWMR-byz-04}, except  line~\ref{MWMR-byz-11}
where $p_i$ writes into the timestamp of $reg_i[i]$  a new epoch. 

The second difference is at lines~\ref{MWMR-byz-14}-\ref{MWMR-byz-16}, 
where the value returned by the read operation is computed.
This value is the one associated with  the greatest epoch known by $p_i$ 
and the greatest sequence number, and where process identities are used
to do tie-breaking (if needed).

\paragraph{Proof} 
Due to page limitation, the proof of the previous construction is given 
in Appendix~\ref{annex-proof-atomic-MWMR}.


\section{Conclusion}
\label{sec:conclusion}
This paper was on the implementation of stabilizing server-based storage 
on top of an asynchronous  message-passing system where up to $t$ servers 
can exhibit a Byzantine behavior. 
A first basic algorithm was represented, which implements a 
single-writer single-reader {\it regular} register stabilizing after 
the first  write invocation. This algorithm tolerates $t<n/8$ if 
communication is asynchronous,  and  $t<n/3$ if it is synchronous.
This algorithm was then extended to obtain a 
{\it practically stabilizing atomic} single-writer single-reader register. 
Finally, the paper presented a generalization allowing any number of processes 
to read and write the practically stabilizing atomic register.

This paper, together with~\cite{BPT15},
is one of the very first to address the construction of a read/write  
register in an asynchronous system where all servers can experience 
transient failures, and some  of them  can behave arbitrarily. 
While the algorithms presented in~\cite{BPT15}, require the ``operation 
quiescence''  assumption, and  build only regular registers, 
(as already noticed in the introduction) our constructions 
are the first that  build a distributed
atomic read/write memory on top of asynchronous servers, which communicate by 
message-passing with the readers and writers processes, can suffer transient 
failures, and where some of them can exhibit a Byzantine behavior.  

\newpage
\setcounter{page}{1}
\pagenumbering{roman}


\newpage

\appendix

\section{SWSR Regular Register in a  Synchronous Communication Setting}
\label{SWSRSS}

This section presents and proves correct an algorithm, which  builds
a stabilizing SWSR regular register, in a synchronous system where up to  
$t<n/3$ servers can commit Byzantine failures. 

As aleady indicated in Section~\ref{sec:synchronous-SWSR}, 
{\it synchronous} means here that  there is an upper bound on message 
transfer delays  on each link connecting a  process 
(reader or writer) and a correct server. Moreover, this 
bound is known by the processes. Hence, both the reader and the writer know
how long they  have to wait for messages from all correct servers, 
and can consequently use timers with appropriate timeout values.

The corresponding algorithm is described in Figure~\ref{algo:SWSRSS-regular-ss},
which is a simple adaptation of the basic algorithm of 
Figure~\ref{algo:SWSR-regular}. The modified lines are suffixed with the 
letter M. 

Due to the link synchrony  property, we have the following.  
When the writer writes a value $x$ to the correct servers (which are at least
$(2t+1)$),  and then starts another  write of a value $y$, 
as it obtains values from  all correct servers, a concurrent read
 obtains at least $(t+1)$ messages carrying  $x$, or at least $(t+1)$ messages 
carrying $y$.  
More generally, if the writer is faster than the reader, it assists the reader 
to find $(2t+1)$ identical non-$\bot$ values, writing the same value at all 
correct servers.
The reader can then read at least $(t+1)$ identical non-$\bot$  values 
in the $helping\_val$ field of the messages it receives from all correct 
servers, and is able to return a correct value.

\begin{figure}[h]
\begin{myalgo}

{\bf operation} $\mathsf{write}$ ($v$) {\bf is}   
                    ~~~~\% issued by the writer $p_w$  \% \\
(\ref{ns-byz-01})  \> ~~~  
${\sf ss\_broadcast}$ {\sc write} $(v)$ ${\sf to}$  all servers;\\

(\ref{ns-byz-02}.M) \>   ~~~ ${\sf wait}$ 
      \big(messages {\sc ack\_write} $(helping\_val)$
                      received from $n$ different servers or time-out\big);\\

(\ref{ns-byz-03}.M) \>  ~~~ {\bf if} ~\=
$\neg$\big($\exists$ $w\neq \bot$ such that $helping\_val=w$ for $(t+1)$ 
                                   of the previous messages\big)\\

(\ref{ns-byz-04}) \>  ~~~\>\>
{\bf then}  
   ${\sf ss\_broadcast}$ {\sc new\_help\_val}$(v)$ ${\sf to}$  all servers\\

(\ref{ns-byz-05}) \>  ~~~ {\bf end if};\\

(\ref{ns-byz-06}) \>  ~~~ ${\sf return}()$.\\~\\

{\bf operation} ${\sf read}$ () {\bf is}   
                    ~~~~\% issued by the reader $p_r$  \% \\

(\ref{ns-byz-07}) \> $\mathit{new\_read_r \leftarrow  {\tt true}}$; \\

(\ref{ns-byz-08}) \> {\bf wh}\={\bf ile}  (${\tt true}$)   {\bf do}\\

(\ref{ns-byz-09}) \>\> 
   ${\sf ss\_broadcast}$ 
{\sc read} $(\mathit{new\_read_r})$ ${\sf to}$    all servers;\\

(\ref{ns-byz-10}) \>\> 
    $\mathit{new\_read_r \leftarrow  {\tt false}}$;  \\

(\ref{ns-byz-11}.M)\>  \> ${\sf wait}$ 
 \big(messages {\sc ack\_read} $(last\_val,helping\_val)$ \\ 
 \>  \>~~~~~~~~~~~~~~~~~~~~~~~~~~                     
received from $n$ different servers or time-out\big);\\

(\ref{ns-byz-12}.M) \>  \> {\bf if} \=
  \big($(t+1)$ of the previous messages
have the same $last\_val$\big)\\
          
(\ref{ns-byz-13}) \>\> \> {\bf then} \=
{\bf let}  $ v$ be this value;  ${\sf return}(v)$

~~~~~ \% the value returned is regular or atomic  \%   \\

(\ref{ns-byz-14}.M) \> \>\> {\bf else} \> 
    {\bf if} \= \big($(t+1)$  of the previous messages
have the same $helping\_val\neq \bot$\big)\\

(\ref{ns-byz-15})\>   \>\>\>\> {\bf then} \=
{\bf let}  $w$  be this value;  ${\sf return}(w)$   
    ~ \% the value returned is atomic \%  \\

(\ref{ns-byz-16})\> \>  \>\>  {\bf end if}\\

(\ref{ns-byz-17}) \>\>  {\bf end if}\\

(\ref{ns-byz-18}) \> {\bf end while}.\\

-----------------------------------------------------------------------------------------------------------\\

{\bf when} {\sc write} $(v)$ {\bf is} 
                 ${\sf ss\_delivered}$  {\bf from} $p_w$ {\bf do}\\

(\ref{ns-byz-19}) \> $last\_val_i\leftarrow v$; \\

(\ref{ns-byz-20}) \> 
${\sf send}$  
{\sc ack\_write} $(helping\_val_i)$ ${\sf to}$ $p_w$.
\\~\\

{\bf when} {\sc new\_help\_val} $(v)$ 
      {\bf is}   ${\sf ss\_delivered}$  {\bf from} $p_w$ {\bf do}\\

(\ref{ns-byz-21}) \>   $helping\_val_i\leftarrow v$.\\~\\

{\bf when}  {\sc read} $(\mathit{new\_read})$
          {\bf is} ${\sf ss\_delivered}$  {\bf from} $p_r$ {\bf do}\\

(\ref{ns-byz-22}) \> {\bf if} $(\mathit{new\_read})$
     {\bf then} $helping\_val_i \leftarrow \bot$ 
     {\bf end if};\\

(\ref{ns-byz-23}) \> 
  ${\sf send}$  {\sc ack\_read} 
     $( last\_val_i,helping\_val_i)$ ${\sf to}$ $p_r$.

\end{myalgo}
\caption{Byzantine-tolerant stabilizing SWSR regular register, 
         (semi-synchronous  links and $t<n/3$)}
\label{algo:SWSRSS-regular-ss}
\end{figure}

The proof is a straightforward adaptation of the proof of 
Section~\ref{sec:basic-algo-proof}, which  takes into account 
the synchrony assumption. It assumes $t<n/3$. 

\begin{lemma}
\label{write-termination-ss-2}
Any invocation of $\mathsf{write}~()$ terminates.
\end{lemma}

\begin{proofL}
Due to the ss-broadcast termination property, the writer cannot block forever
when it invokes ${\sf ss\_broadcast}()$ at line~\ref{ns-byz-01} or 
line~\ref{ns-byz-04}. As far the wait statement of line~\ref{ns-byz-02} 
is concerned, we have the following. Due to the ss-broadcast eventual delivery 
property, at least $(n-t)$ non-Byzantine servers ss-deliver 
the message  {\sc write}$()$ ss-broadcast by the writer,  and send it by 
return the acknowledgment message {\sc ack\_write}$()$, which concludes 
the proof of the lemma.  
\renewcommand{\toto}{write-termination-ss-2}
\end{proofL}

\begin{lemma}
\label{read-termination-ss-2}
Any invocation of $\mathsf{read}~()$ terminates. 
\end{lemma}

\begin{proofL}
Using the same reasoning as in Lemma~\ref{write-termination}, 
it follows that the reader cannot block forever in the wait statement of 
line~\ref{ns-byz-11}. So, the proof consists in showing that the predicate of 
line~\ref{ns-byz-12}, or the one of~\ref{ns-byz-14}, becomes eventually true. 
The rest of the proof is by contradiction.  $R$ being the first invocation 
of $\mathsf{read}~()$ that does not terminate, let us consider an 
execution of the loop body after time $\tau_{stab}$. 

{\sl Claim} C. At the time at which a write that started after $\tau_{no\_tr}$ 
terminates, there are (a) at least $(n-t=(2t+1))$ correct servers $s_i$ 
such that  $last\_val_i=v$, and  (b) at least $(t+1)$ correct servers $s_j$ 
such that $helping\_val_j=w\neq\bot$.\\
{\sl Proof of the claim}.
It follows from the synchronized delivery  property of the
ss-broadcast of the message {\sc write}$()$, and the fact that
no correct server suffers transient failures after $\tau_{no\_tr}$,
that, when a write that started after $\tau_{no\_tr}$ terminates,
there are at least $(n-t)$ correct servers $s_i$ such that $last\_val_i=v$.
Moreover, if the predicate of line~\ref{ns-byz-03} is true, it follows from
(a) the synchronized delivery property of the ss-broadcast of the message
{\sc new\_help\_val}$()$ (line~\ref{ns-byz-04}), and (b) the fact that 
$n-t>t+1$, that at least $(t+1)$ correct servers $s_j$ are such that
$helping\_val_j=w\neq\bot$.
If predicate of line~\ref{ns-byz-03} is false, there are $((2t+1))$
servers that sent {\sc ack\_write}$(w)$ where $w\neq \bot$
(line~\ref{ns-byz-20}),
from which we conclude that there are at least $(t+1)$ with 
$helping\_val_j=w\neq\bot$. 
{\sl End of the proof of the claim} C.

Let us consider the last  write that terminated before $R$ 
started,  and let us assume it wrote $x$.  Due to part (a) of Claim C,
just after this write terminated, all the $(n-t)$ correct servers $s_i$ 
are such that $last\_val_i=x$.  If no write is concurrent with $R$, as $R$
receives messages {\sc ack\_read}$(last\_val,-)$ from $(n-t)$ 
correct servers at line~\ref{ns-byz-11}, 
it follows that $R$ terminates at lines\ref{ns-byz-12}-\ref{ns-byz-13}. 

Let us now assume that there is exactly one write that is 
concurrent with $R$, and  let $y$ be the value it writes.
Due to the synchronized delivery property of ss-broadcast, $R$ first resets 
to $\bot$ the variables $helping\_val_i$ of all $n-t$ 
correct servers $s_i$ (lines~\ref{ns-byz-07},~\ref{ns-byz-09}, 
and~\ref{ns-byz-22}), and then receives (line~\ref{ns-byz-11}) messages 
{\sc ack\_read}$(last\_val,-)$ from all the correct servers.
We show that at least $(t+1)$ of these messages carry either $x$ or $y$, 
from which $R$ terminates at lines~\ref{ns-byz-12}-\ref{ns-byz-13}.  
Due to part (a) of Claim C, there were at least $n-t \geq (2t+1)$ correct 
servers $s_i$ such that $last\_val_i=x$  when the write of $x$ finished.
Let $Q$ be this set of servers.
$R$ receives messages {\sc ack\_read}$(last\_val,-)$  from all the  
$((2t+1))$ correct servers in $Q$.
Due to the operation ${\sf write}(y)$ (concurrent with $R$), variables
$last\_val_i$ of some of these servers may have been updated to the value $y$.
Hence, some of the previous $((2t+1))$ messages {\sc ack\_read}$(last\_val,-)$
received by $R$ carry $x$, while others carry $y$. Hence, at least $(t+1)$
of them carry either $x$ or $y$, and $R$ terminates
at lines~\ref{ns-byz-12}-\ref{ns-byz-13}.

Let us finally consider the case where there are more than one write 
concurrent with $R$. When $R$ terminates its  invocation of
${\sf ss\_broadcast}$ {\sc read}$({\tt true})$  
(there is only one such invocation per read, line~\ref{ns-byz-09}), 
the local variables $helping\_val_i$ of $(n-t)$ correct servers are equal 
to $\bot$. Let $Q'$ be this set of servers. 
(The proof of this statement is the same as the proof appearing in the
first part of  claim C.) 
Hence, when this ss-broadcast terminated,  the messages 
{\sc ack\_read}$(-,helping\_val_i)$  sent by each  server $s_i\in Q'$
(line~\ref{ns-byz-23}),  is such that $helping\_val_i=\bot$. 
Let us consider the first write (e.g., ${\sf write}(z)$)
that occurs after the servers $s_i\in Q'$
have set $helping\_val_i$ to $\bot$. This write receives $(n-t)$
messages {\sc ack\_write} $(helping\_val)$, and at least
$((2t+1))$ of them are from servers in $Q'$, 
and carry consequently  $helping\_val=\bot$.  Hence the 
predicate of line~\ref{ns-byz-03} is satisfied, and the writer issues
${\sf ss\_broadcast}$~{\sc new\_help\_val}$(z)$. 
If later (i.e.,  after the invocation of  ${\sf write}(z)$ terminated),
there are other invocations of  ${\sf write}()$ concurrent with $R$, 
none of them will execute line~\ref{ns-byz-04}. This is due to the fact that 
$R$ does not reset  the variables $helping\_val_i$ to $\bot$, 
and the $(n-t)$ messages {\sc ack\_write}~$(helping\_val)$ sent by the servers
at line~\ref{ns-byz-20} are such that at most $t$ are from Byzantine servers, 
and at least $((2t+1))$ carry $z$, from which 
follows that there is a finite time $\tau_R$ after which the variables 
$helping\_val_i$ of the correct servers are no longer modified. Let us 
finally consider the first invocation of ${\sf ss\_broadcast}~(new\_read_r)$
issued by $R$ after  $\tau_R$,  such that  $new\_read_r={\tt false}$.
It follows from the previous discussion that, among the 
$(n-t)$ messages {\sc ack\_read}$(-,helping\_val)$ received by $R$,
at most $t$ (the ones from Byzantine servers) carry arbitrary values, 
and at least $(n-t)$ carry the value $z$.  When this occurs, 
$R$ terminates at  lines~\ref{ns-byz-14}-\ref{ns-byz-15}.
\renewcommand{\toto}{read-termination-ss-2}
\end{proofL}

\begin{lemma}
\label{label:eventual-regularity-ss-2}
Let $t<n/3$.
There is a finite time  $\tau_{stab}>\tau_{1w}$
after which each read invocation $R$ returns a value  $v$  that 
was written by a write operation $W$, which is 
(a) the last write operation executed before $R$, or 
(b) a write operation concurrent with $R$. 
\end{lemma}

\begin{proofL}
Let us assume that a read operation $R$
 returns $z$,  a value  different from
the value $v$ of the  last completed write prior to $R$,  
and from any  value $u$ of a concurrent write. 
Let us consider the first write concurrent with $R$. For $R$ to return
$z$, the reader must receive $(t+1)$ messages
{\sc ack\_read}$(z,-)$ or $(t+1)$ messages {\sc ack\_read}$(-,z)$.
However, immediately following the termination of the write of $v$ there 
were $(n-t)$ correct servers $s_i$ 
with $last\_val_i=v$. Thus,  following the termination of the write of $v$,
and until the termination of the next write of some value $u$,
the reader cannot receive 
$(t+1)$ values for a value  $z$ different from $v$ and $u$. 
The above argument holds for the second concurrent write, where we start 
with $(n-t)$ values of $u$, and so on and so forth.
\renewcommand{\toto}{label:eventual-regularity-ss-2}
\end{proofL}

\begin{theorem-repeat}{theorem:regular-register-ss-2}
Let $t<n/3$. The algorithm described in Figure~\ref{algo:SWSRSS-regular-ss}
implements a stabilizing regular register in the presence of at 
most $t$ Byzantine servers.  
\end{theorem-repeat}

\begin{proofT}
The proof follows from Lemma~\ref{write-termination-ss-2}, 
Lemma~\ref{read-termination-ss-2}, and 
Lemma~\ref{label:eventual-regularity-ss-2}.
\renewcommand{\toto}{theorem:regular-register-ss-2}
\end{proofT}

\section{Proof of the Stabilizing SWSR Atomic Register
(Section~\ref{sec:practically-atomic})}
\label{annex-proof-practically-atomic-SWSR}

\begin{lemma}\label{lem:SWSRAWTerm}
Any invocation of a ${\sf prac\_at\_write}()$ operation terminates.
\end{lemma}

\begin{proofL}
Let us suppose by contradiction that there exists a ${\sf prac\_at\_write}()$ 
operation $op_w$ invoked by the writer $p_w$ and that $op_w$ does not terminate.
If such operation does not terminate, it means that $p_w$ never executes 
line \ref{ns-byz-06} in Figure \ref{algo:SWSR-atomic}.
Let us note that, due to the ss-broadcast termination property, 
$p_w$ cannot be blocked while sending messages. 
Thus, the only point where $p_w$ can be blocked is executing 
line \ref{ns-byz-02} in Figure \ref{algo:SWSR-atomic} while waiting for 
the delivery of {\sc ack\_write}$()$ messages.
An {\sc ack\_write}$()$ message is sent by a server when it delivers a 
{\sc write}$(ts, v)$ message (line \ref{ns-byz-20}, 
 Figure \ref{algo:SWSR-atomic}) that is in turn sent by $p_w$ at the 
beginning of the ${\sf prac\_at\_write}()$ operation (line \ref{ns-byz-01}M,  
Figure \ref{algo:SWSR-atomic}).
Due to the eventual delivery property of ss-broadcast, we have that
 eventually $n-t$ correct servers will deliver the {\sc write}$()$
 message sent by $p_w$ and will send back an {\sc ack\_write}$()$ message.
Thus, considering that links connecting each server to the writer is
 FIFO reliable, we have that  $p_w$ will eventually deliver at least $n-t$ 
{\sc ack\_write}$()$ messages. 
Therefore, we have a contradiction and the claim follows.
\renewcommand{\toto}{lem:SWSRAWTerm}
\end{proofL}

\begin{lemma}\label{lem:stateAfterWrite}
Let $op_w$ be a ${\sf prac\_at\_write}(v)$ operation invoked by the writer 
$p_w$ at some time $t_S(op_w) \ge \tau_{no\_tr}$, let $wts$ be the sequence 
number associated to the operation and let $t_E(op_w)$ be the time at
 which $op_w$ terminates. At time $t_E(op_w)$ there exist at least $(n-2t)$ 
correct servers that store locally in their $last\_val_i$ variable the pair 
$\langle v, wts \rangle$.
\end{lemma}

\begin{proofL}
Due to Lemma \ref{lem:SWSRAWTerm}, we have that time $t_E(op_w)$ exists. 
Let us now show that at that time, at least $(n-2t)$ correct servers store 
the pair $\langle v, wts \rangle$.
The writer $p_w$ returns from the ${\sf prac\_at\_write}(v)$ operation 
only after it is unlocked from the ${\sf wait}$ statement in 
line \ref{ns-byz-02}.
If $p_w$ is unblocked, it means that it delivered at least $(n-t)$ 
{\sc ack\_write}$()$ messages from $n-t$ different servers.
An {\sc ack\_write}$()$ message is sent by a server $s_i$ when it delivers 
a {\sc write}$(v)$ message and just after it updated its local copy of the 
register with the value and the sequence number contained in the 
{\sc write}$(v)$ message (line \ref{ns-byz-19},
 Figure \ref{algo:SWSR-atomic}). Let us denote as $\tau_{update}$ such a time.
Considering that (i) both ss-broadcast and the FIFO 
link involved in such a message pattern do not create messages, 
(ii) the value and the sequence number are communicated to $s_i$ directly 
from the writer, (iii) among the $(n-t)$messages  {\sc ack\_write}$()$ 
received by $p_w$, at most $t$ are from  Byzantine servers,  and 
(iv) $\tau_{update} < t_E(op_w)$, the claim follows.
\renewcommand{\toto}{lem:stateAfterWrite}
\end{proofL}

\begin{lemma}\label{lem:stateHVAfterWrite}
Let $op_w$ be a ${\sf prac\_at\_write}(v)$ operation invoked by the writer 
$p_w$ at some time $t_S(op_w) \ge \tau_{no\_tr}$, let $wts$ be the sequence 
number associated to $op_w$ and let $t_E(op_w)$ be the time at which $op_w$
 terminates. At time $t_E(op_w)$ there exist at least $(4t+1)$ correct servers 
that store locally in their $helping\_val_i$ variable the same pair 
$\langle v', ts \rangle$.
\end{lemma}

\begin{proofL}
Due to Lemma \ref{lem:SWSRAWTerm}, we have that time $t_E(op_w)$ exists. 
Let us now show that at that time, at least $(n-2t)$ correct servers store 
the same pair $\langle v', ts\rangle$.
The writer $p_w$ returns from the ${\sf prac\_at\_write}(v)$ operation only 
after it is unblocked from the ${\sf wait}$ statement in line \ref{ns-byz-02}.
If $p_w$ is unblocked, it means that it delivered at least $(n-t)$ 
{\sc ack\_write}$(hv)$ messages from $(n-t)$ different servers. 
Thus, $p_w$ received at least $(n-t)$ helping values, stored locally 
at the servers, from $(n-t)$ different servers.
Let $t_{del}$ be the time at which $p_w$ is unblocked from the ${\sf wait}$ 
statement in line \ref{ns-byz-02} and evaluates the condition in line 
\ref{ns-byz-03}.
Two cases can happen: the condition at line~\ref{ns-byz-03} is 
(i) $\mathit{\tt true}$.  or (ii) $\mathit{\tt false}$. 
\begin{itemize}
\vspace{-0.1cm}
\item {\sl Case 1: The condition in line \ref{ns-byz-03} is 
${\mathit{\tt true}}$}. 
In this case, it means that among the $(n-t)$ received helping values, 
there not exists a value $w \neq \bot$ occurring a majority of time. 
This means that helping values stored at each server $s_i$ during the current 
${\sf prac\_at\_write}()$ operation are corrupted values and need to be cleaned.
Thus, at time $t_{del}$, the writer $p_w$ broadcasts a 
{\sc new\_help\_val}$(wts, v)$ message that will trigger the update of
 the $helping\_val_i$ variable (line \ref{ns-byz-21}).
Considering that ss-broadcast 
(i) does not modify the content of messages, 
(ii) guarantees that at least $(n-2t)$ correct servers deliver the message 
before the end of its invocation, and 
(iii) $p_w$ returns form the ${\sf prac\_at\_write}(v)$ operation only 
after the termination of the ss-broadcast, it follows that at least $(n-2t)$ 
correct servers stored the same pair $\langle v, wts \rangle$ in their 
$helping\_val_i$ local variable before the end of the operation. 
As $n>8t$, the claim follows.
\vspace{-0.2cm}
\item {\sl Case 2: The condition in line \ref{ns-byz-03} is 
${\mathit{\tt false}}$}.
 In this case, the claim directly follows as the writer found $(4t+1)$ same
 values.
\end{itemize}
\vspace{-0.6cm}
\renewcommand{\toto}{lem:stateHVAfterWrite}
\end{proofL}

\begin{lemma}\label{lem:SWSRARTerm}
Any invocation of a ${\sf prac\_at\_read}()$ operation terminates.
\end{lemma}

\begin{proofL}
Let us suppose by contradiction that there exists a ${\sf prac\_at\_read}()$ 
operation $op_r$ invoked by the reader $p_r$ and that $op_r$ does not terminate.
If such operation does not terminate, it means that $p_r$ never executes 
line \ref{ns-byz-13}M2 or line \ref{ns-byz-13}M3 or line~\ref{ns-byz-15}M 
in Figure \ref{algo:SWSR-atomic}.
Let us note that, due to the  ss-broadcast termination property, 
$p_r$ cannot be blocked while sending messages. 
Thus, the only points where $p_r$ can be blocked is (i) while executing 
line \ref{ns-byz-11} in Figure \ref{algo:SWSR-atomic} keep waiting for 
the delivery of {\sc ack\_read}$()$ messages or (ii) cycling for ever as 
the set of {\sc ack\_read}$()$ messages received by clients never contains 
two values $x$ and $y$ such that $x$ is the last value reported by at 
least $(2t+1)$ servers or $y$ is the helping value reported by at least 
$(2t+1)$ servers.\\

\noindent {\sl Case 1: The reader remains blocked while executing 
line \ref{ns-byz-11} in Figure \ref{algo:SWSR-atomic}.} 
If the reader is blocked while executing line \ref{ns-byz-11} in
 Figure \ref{algo:SWSR-atomic}, it means that it never delivers at least 
$(n-t)$ {\sc ack\_read}$()$ messages from servers.
An {\sc ack\_read}$()$ message is sent by a server when it delivers a 
{\sc read}$()$ message (line \ref{ns-byz-23},  Figure \ref{algo:SWSR-atomic}) 
that is in turn sent by $p_r$ at the beginning of the ${\sf read}()$ 
operation (line \ref{ns-byz-09},  Figure \ref{algo:SWSR-atomic}).
Due to the eventual delivery property of ss-broadcast, we have that 
eventually $(n-t)$ correct servers will deliver the {\sc read}$()$ message 
sent by $p_r$ and will eventually send back a {\sc ack\_read}$()$ message.
Thus, considering that links connecting each server to the writer is FIFO 
reliable, we have that  $p_r$ will eventually deliver at least $(n-t)$ 
{\sc ack\_read}$()$ messages. Therefore, we have a contradiction and 
this case can never happen.\\

\noindent{\sl Case 2: The reader never collects $(2t+1)$ copies of 
the same last value 
or it never collects $(2t+1)$  copies of of the same helping value.} 
Let us note that last 
values and helping values are sent from a server $s_i$ trough an 
{\sc ack\_read}$()$ message when it delivers a {\sc read}$()$ message 
(line \ref{ns-byz-23}, Figure \ref{algo:SWSR-atomic}).\\
Thus, if the servers is not able to find $(2t+1)$ same last values or 
$(2t+1)$ same helping values it means that there always exists $(n-t)$ 
servers answering with different values.
Note that each server $s_i$ updates its $last\_val_i$ variable while 
delivering a {\sc write}$()$ message sent by the writer and it updates 
its $helping\_val_i$ variable either during a write using values provided 
by the writer or during a read resetting such value to $\bot$.
Considering that, by assumption, there exists a ${\sf prac\_at\_write}()$ 
operation issued after time $\tau_{no\_tr}$ we have that, due to 
Lemma \ref{lem:stateAfterWrite} and Lemma \ref{lem:stateHVAfterWrite} 
there exists a time $\tau> \tau_{no\_tr}$ at which the write terminates and 
such that at least $(n-2t)$ correct servers store the same last value and 
such that at least $(4t+1)$ correct server stores the same helping value.
Let us show now that the ${\sf prac\_at\_read}()$ operation $op_r$ eventually 
terminates after time $\tau$. 
Let us consider the first {\sc read}$()$ message $m$ broadcast by $p_r$ 
after time $\tau$.
Two further cases may happen: (2.1) $m$ is the first message sent by $p_r$ 
in the while loop (i.e., $m$ is a {\sc read}$(\mathit{\tt true})$  message 
(line \ref{ns-byz-09}),  or (2.2) $m$ is the $\alpha$-th message sent by $p_r$ 
in the while loop, with $\alpha>1$ (i.e., $m$ is a 
{\sc read}$((\mathit{\tt false})$ message (line~\ref{ns-byz-09}).
\begin{itemize}
\vspace{-0.1cm}
\item {\sl Case 2.1.} if $m$ is a {\sc read}$(\mathit{\tt true})$ message, 
it will trigger the update of the $helping\_val_i$ variable to $\bot$ 
at any correct server $s_i$ that will deliver it. Due to the Synchronized 
Delivery property of the ${\sf ss-broadcast}$ primitive, we have that at 
least $(n-2t)$ correct servers will update their $helping\_val_i$ variable.
Considering that, at time $\tau$, we have $(n-2t)$ correct servers storing the
 same last value and considering that we have only one reader $p_r$, 
it follows that such values can be modified concurrently with the broadcast 
only by the writer. So, if the writer is not going to modify such values, 
servers will answer to the broadcast by sending back last values stored at 
time $\tau$ and the helping values just updated. Considering that messages 
are not altered by the network, the reader will receive at least $n-3t$ 
same last values and at least $n-3t$ same helping values. 
Thus, evaluating the condition in line \ref{ns-byz-12}, the reader 
will find it true and it will terminate the operation either 
executing line~\ref{ns-byz-13}M2 or line~\ref{ns-byz-13}M3.\\
Contrarily, if the writer is going to update the $last\_val_i$ variables
 due to a concurrent write, the reader will find the condition in 
line  \ref{ns-byz-12} false as well as the condition in line  \ref{ns-byz-14}.
Note that such concurrent write will be acknowledged by servers with at 
least $n-3t$  helping values equal to $\bot$. 
This will entail the update of the
 $helping\_val_i$ variables with the value concurrently written.
As a consequence, in the next iteration of the while loop, due to 
lemma \ref{lem:stateHVAfterWrite}, there will exist at least $(4t+1)$
 correct servers with the same helping value different from $\bot$. 
Thus, at least $(2t+1)$ will acknowledge the next {\sc read}$()$ message making 
the condition in line \ref{ns-byz-14} true and letting the operation terminate.
\vspace{-0.2cm}
\item {\sl Case 2.2.} If $m$ is a {\sc read}$(\mathit{\tt false})$ message, 
it will just be acknowledge by servers with the current values stored locally 
in their $last\_val_i$ variable and in their $helping\_val_i$ variable.
Considering that, at time $t$, we have $(n-2t)$ correct servers storing 
the same last value, we have at least $(4t+1)$ correct servers storing 
the same helping values and considering that we have only one reader $p_r$, 
it follows that such values can be modified concurrently with the broadcast
 only by the writer. 
Depending on the value stored by the $(4t+1)$ correct servers (i.e., $\bot$ 
or a different one) we fall down in the previous case or we have that the 
reader will find the condition in line \ref{ns-byz-14} immediately true.
However, 
in both case we have the termination of the operation and the claim follows.
\end{itemize}
\vspace{-0.4cm}
\renewcommand{\toto}{lem:SWSRARTerm}
\end{proofL}

\begin{lemma}\label{lem:TOafterStab}
Let $\tau_{no\_tr}$ be the time after which no more transient failures happen. 
Let $op_{w_1}$ be the first ${\sf prac\_at\_write}()$ operation issued after 
$\tau_{no\_tr}$ and let $\tau_{1w} > \tau_{no\_tr}$ be the time at which 
$op_{w_1}$ terminates.
Let $S_W$ be the sequence of ${\sf prac\_at\_write}()$ operations issued 
by $p_w$ and let $S_W|op_{w_1}$ be the sub-sequence of $S_W$ starting with 
$op_{w_1}$.
For each $op_{w_i} \in S_W|op_{w_1}$, let $sn_i$ be the sequence number 
associated to the operation.
For each pair $op_{w_i}$, $op_{w_{i+1}}$ of adjacent write in $S_W|op_{w_1}$ 
we have that $sn_i \prec sn_{i+1}$.
\end{lemma}

\begin{proofL}
The claim simply follows by the definition of the precedence relation 
$>_{cd}$ considering that after time $\tau_{no\_tr}$ the sequence number 
is generated only by the unique writer by incrementing the previous one.
\renewcommand{\toto}{lem:TOafterStab}
\end{proofL}

\begin{lemma}\label{lem:evValidityAtomic}
Let $t<n/8$. There is a finite time  $\tau_{stab}>\tau_{1w}$ after which 
each ${\sf prac\_at\_read}()$ operation $op_r$ returns a value $v$ that
 was written by a write operation $op_w$, which is (a) the last 
${\sf prac\_at\_write}()$ operation executed before $op_r$, or 
(b) a ${\sf prac\_at\_write}()$ operation concurrent with $op_r$. 
\end{lemma}

\begin{proofL}
Due to Lemma \ref{lem:SWSRARTerm}, we have that eventually each
 ${\sf prac\_at\_read}()$ operation terminates. Let us show in the 
following that there exists a time $\tau_{stab}$ after which, 
each ${\sf prac\_at\_read}()$ operation terminates returning a valid value 
(i.e., the last value written or a value concurrently written).
Without loss of generality, let us consider only ${\sf prac\_at\_read}()$ 
operations starting after time $\tau_{1w}$ (i.e., considering only 
${\sf prac\_at\_read}()$ operations following the end of the first 
completed write in the stability period).

Let $op_w$ be the first ${\sf prac\_at\_write}(v)$ operation terminated 
after $\tau_{stab}$ and let $x$ be the sequence number associated to such
 operation and terminated at time $\tau_{1w}$.
Let us consider a ${\sf prac\_at\_read}()$ operation $op_r$ issued at some
time after $\tau_{1w}$. 
When executing $op_r$, the reader $p_r$ sends a {\sc read}$()$ message 
to all servers that will answer by sending back their pair 
$\langle last\_val_i, helping\_val_i \rangle$ (line \ref{ns-byz-23}, 
Figure \ref{algo:SWSR-atomic}).
Note that, due to Lemma \ref{lem:stateAfterWrite}, at time $\tau_{1w}$, 
there exist at least $(n-2t)$ correct servers storing the same 
pair $\langle v, x \rangle$ in their $last\_val_i$ variable and, 
due to Lemma \ref{lem:stateHVAfterWrite}, at time $\tau_{1w}$, 
there exist at least $(4t+1)$ correct servers storing the 
same pair $\langle v', x' \rangle$ in their $helping\_val_i$ local variable.

If there is no concurrent ${\sf prac\_at\_write}()$ operation, 
it means that servers will answer by sending back the value 
$\langle v, x \rangle$ and the pair $\langle v', x' \rangle$.
In order to select a value to return, the reader waits for $(n-t)$ messages, 
that $t$ answers may arrive from Byzantine servers and $t$ may arrive from 
servers that are not yet updated, we have that only $n-3t$ values are 
guaranteed to arrive from correct and updated servers.
Considering that $n>8t$ we hate that at least $5t+1$ messages arrives 
from correct and updates servers. Thus, evaluating the condition in 
line \ref{ns-byz-12}, Figure \ref{algo:SWSR-atomic}, $p_r$ will find 
it true and will check whether $x$ is smaller or greater than its 
current local sequence number.
Two cases may happen: (1) $pwsn >_{cd} x$ or (2) $x \geq_{cd} pwsn$.
\begin{itemize}
\vspace{-0.2cm}
\item {\sl Case 1:  $pwsn >_{cd} x$.} In this case the reader executes 
line \ref{ns-byz-13}M3 returning the value locally stored that can be a 
corrupted one. 
Let us remark that since $op_r$ is the first read executed after the 
stabilization, we may have that executing lines N2 -N7, $p_r$ collects
 helping values that are still corrupted and set the its local sequence 
number to a value that is corrupted.
However, this happen only this time as from this time on, the only process 
that will generate sequence number for write operation is the writer.
Considering that the such sequence number is generated by incrementing each 
time the old one (see Lemma~\ref{lem:TOafterStab}), we have that 
such a scenario may happen a finite number of time. 
Thus, eventually the writer will use a sequence number that is greater
 equal than the current one and we will have that eventually a read returns 
a valid value.
\vspace{-0.2cm}
\item {\sl Case 2: $x \geq_{cd} pwsn$.} 
In this case the reader executes line \ref{ns-byz-13}M2 returning the last 
written value and the claim follows.
\end{itemize}
Let us note that, due to the enforcement of the helping value by the writer, 
we obtain, in case of concurrent writes, the scenario described 
so far,  and the claim follows.
\renewcommand{\toto}{lem:evValidityAtomic}
\end{proofL}

\begin{lemma}\label{lem:evAtomocity}
Let $t<n/8$. 
There is a finite time  $\tau_{stab}>\tau_{1w}$ after which any 
${\sf prac\_at\_read}()$ having less than $2^{63}+1$ ${\sf prac\_at\_write}()$ 
concurrent operations  returns a regular value and no two invocations of 
${\sf prac\_at_read}()$ return new/old inverted values. 
\end{lemma}

\begin{proofL}
Eventual validity follows from \ref{lem:evValidityAtomic} thus, 
in the following, we just need to prove that there exists a time 
$\tau_{stab}>\tau_{1w}$ after which no new/old inversion happens.
Let us suppose by contradiction that there exists two ${\sf prac\_at\_read}()$ 
operations $op_r1$ and $op_r2$ such that $op_r1$ happens before $op_r2$ $op_r1$
 returns a value $v_1$ and $op_r2$ returns a value $v_2$ and
 ${\sf prac\_at\_write}(v_2)$ happens before ${\sf prac\_at\_write}(v_1)$.
If ${\sf prac\_at\_write}(v_2)$ happens before ${\sf prac\_at\_write}(v_1)$ 
it means that $sn_1 >_{cd} sn_2$.
Note that, if $op_r1$ returned value $v_1$, it means that $p_r$ executed
 line \ref{ns-byz-13}M3 ore line \ref{ns-byz-15}M. However, in both cases, 
before returning $v_1$, $p_r$ updated its current local sequence number 
to $sn_1$.
Thus, executing $op_r2$, evaluating the condition in line \ref{ns-byz-13}M2, 
$p_r$ will find it false and will execute line \ref{ns-byz-13}M3 returning 
$v_1$ and we have a contradiction.
 
Note that, the local sequence number can be reset to a value smaller than
 $sn_1$ only if, executing line N6, $p_r$ found the condition true. 
However, this happen if and only if the writer sequence number wrapped 
around as there are more than $2^{63}+1$ concurrent operations and the 
claim follows.
\renewcommand{\toto}{lem:evAtomocity}
\end{proofL}

\begin{theorem}\label{theo:evAtomocity}
Let $t<n/8$.
The algorithm described in Figure~{\em \ref{algo:SWSR-atomic}}  
implements a Byzantine-tolerant practically stabilizing SWSR atomic register. 
\end{theorem}

\begin{proofT}
The proof follows from Lemmas~\ref{lem:SWSRAWTerm}-\ref{lem:evAtomocity}. 
\renewcommand{\toto}{theo:evAtomocity}
\end{proofT}

The synchronous link version has an analogous proof for $t<n/3$.  

\section{Proof of the Stabilizing MWMR Atomic Register
(Section~\ref{sec:single-to-multi})}
\label{annex-proof-atomic-MWMR}

The proof of both the next lemmas is straightforward, as the code of 
$\mathsf{mwmr\_write}~()$ and $\mathsf{mwmr\_read}~()$ is sequential.

\begin{lemma}
\label{write-termination-mwmr}
 Any invocation of $\mathsf{mwmr\_write}~()$ terminates.
\end{lemma}

 \begin{lemma}
\label{read-termination-mwmr}
Any invocation of $\mathsf{mwmr\_read}~()$ terminates.
\end{lemma}

\begin{definition}[Total order relation $\succ_{to}$]
Let $W_i$,  timestamped $(epoch_i, seq_i)$, be any write issued 
any process $p_i$, and  $W_j$ timestamped $(epoch_j, seq_j)$ be 
any write issued by any process $p_j$.  
$W_j \succ_{to} W_i$ iff $(epoch_j \succ epoch_i)$ $\vee$ $((epoch_j = epoch_i) 
\wedge (seq_j>seq_i))$ $\vee$ $((epoch_j = epoch_i) \wedge (seq_j=seq_i) 
\wedge (j > i))$. Moreover,  
$W_j\succeq_{to}W_i$ $\equiv$ $\big((W_j \succ_{to} W_i)~ \vee~ (W_i=W_j)\big)$. 
\end{definition}

\begin{lemma}[Total order on writes]  
\label{label:eventual-tow}
 There is a finite time  $\tau$ that follows either a non concurrent 
write or non concurrent read, such that  all write operations 
invoked after  $\tau$ are totally ordered. 
\end{lemma}
\begin{proofL}
First notice that non concurrent write or read enforces the existence of 
a greatest epoch which all subsequent read and write identify. 
Let $S$  be the set of writes on the register happened after $\tau$. 
We will prove in the following that $\succeq_{to}$ is a total order on S:
\begin{itemize}
\vspace{-0.2cm}
\item $\succeq_{to}$ reflexivity,  $W_i$ $\succeq_{to}$ $W_i$, follows 
directly from the definition.
\vspace{-0.2cm}
 \item $\succeq_{to}$ antisymmetry, 
 $(W_i \succeq_{to} W_j) \wedge (W_j \succeq_{to} W_i)$ implies $W_i=W_j$. 
Since $W_i$ and $W_j$ happen after $\tau$ it follows that the above 
relations reduce to $(i \geq j) \wedge (j \geq i)$. Hence $i=j$ and $W_i=Wj$.  
\vspace{-0.2cm}
 \item $\succeq_{to}$ transitivity, 
$(W_i \succeq_{to} W_j) \wedge (W_j \succeq_{to} W_k)$ implies  
$(W_i \succeq_{to} W_k)$. This follows directly from the definition 
and the fact the invocation time is after $\tau$.
\vspace{-0.2cm}
\item $\succeq_{to}$ comparability, for any $W_i$ and $W_j$ in $S$, 
$(W_i \succeq_{to} W_j)$ or $(W_j \succeq_{to} W_i)$. This follows directly 
from the definition and the fact that the writes happen after the $\tau$.
 \end{itemize}
\vspace{-0.5cm}
\renewcommand{\toto}{label:eventual-tow}
\end{proofL} 

\begin{lemma}[Regularity]
\label{label:eventual-regularity-mwmr}
 There is a finite time that follows either a non-concurrent write 
or non-concurrent read,
after which each read invocation $R$ returns a regular value.
\end{lemma}
 
\begin{proofL}
In the following we prove that value  $v$  returned by $R$ is the value that 
was written by a write operation $W$, which is 
(a) the last write operation executed before $R$, or 
(b) a write operation concurrent with $R$. 
Following Lemma \ref{label:eventual-tow} there is a time $\tau$ 
such that all writes invoked after $\tau$ are totally ordered. 
Let $R$ be a read operation that happens after $\tau$.
Let $W$ be the last writer in that order that modified the register after 
$\tau$  before $R$ started. $W$ either happened before $R$ or is
 concurrent with $R$.   
The reader $R$ reads first all the SWMR registers and stores their 
values in the vector $reg$. Let $k$ be the index of the SWMR register 
corresponding to $W$.  Since, $W$ is the last writer on the register 
according to $\succeq_{to}$ it follows that 
$reg[k].epoch={\sf  max\_epoch}(reg_i[1..m])$ 
and $reg[k].seq \geq reg[j].seq, 
                   \forall j, reg[j].epoch={\sf  max\_epoch}(reg_i[1..m])$ 
(lines~\ref{MWMR-byz-09}-\ref{MWMR-byz-10}, Figure \ref{algo:MWMR-atomic}) 
and $k$ is the minimal with this property. It follows that $R$ returns 
$reg[k].v$ which is the value written by $W$.  
\renewcommand{\toto}{label:eventual-regularity-mwmr}
\end{proofL}

\begin{lemma}[No new/old inversion]
\label{lemma-no-new-old-SMWMR}
There is a finite time  $\tau$  that follows either a non-concurrent write 
or non-concurrent read,
after which read invocations do not return new/old inverted values.
\end{lemma}
\begin{proofL}
Following Lemma \ref{label:eventual-tow} there is a time $\tau$ 
such that all writes invoked after $\tau$ are totally ordered. 
Let $R_1$ and $R_2$ be two read operations that happen after $\tau$ 
and let $W_1$ and $W_2$ that also happen after $\tau$. Assume also 
that $R_1$ happens before $R_2$, $W_1$ happens before $W_2$ (and no other 
write happens after $W_1$ and before $W_2$), $R_1$ is concurrent with $W_1$ 
and $W_2$ and $R_2$ is concurrent with $W_2$. Assume a new/old inversion
 on $R_1$ and $R_2$. That is, $R_1$ returns the value written by $W_2$ and 
$R_2$ returns the value written by $W_1$.

Let $m1$ be the index in $reg_{R_1}$  that stores the state of the register
 modified by $W_2$.
Let $m2$ be the index in $reg_{R_2}$  that stores the state of the register
 modified by $W_1$.
Since $W_1$ happens before $W_2$ then 
$reg_{R_1}[m1].epoch \succ reg_{R_2}[m2].epoch$ or $reg_{R_1}[m1].epoch
     = reg_{R_2}[m2].epoch$ and $reg_{R_1}[m1].seq > reg_{R_2}[m2].seq$. 
It follows that  we have $reg_{R_2}[m1].epoch \succ reg_{R_2}[m2].epoch$,
 or  $reg_{R_2}[m1].epoch = reg_{R_2}[m2].epoch$ and 
$reg_{R_2}[m1].seq > reg_{R_2}[m2].seq$. 
Hence, $R_2$ has to return the value stored at the index $m1$ which 
corresponds to the value written by $W_2$. This contradicts
the new/old inversion assumption. 
\renewcommand{\toto}{lemma-no-new-old-SMWMR}
\end{proofL}

\begin{theorem}\label{theorem:evAtomocity}
Let $t<n/8$ for the asynchronous version and $t<n/3$ for the link synchronous 
version. The algorithm described in Figure~{\em \ref{algo:MWMR-atomic}}  
implements a Byzantine-tolerant  stabilizing MWMR atomic register. 
\end{theorem}

\begin{proofT}
The proof follows from 
Lemmas~\ref{write-termination-mwmr}-\ref{lemma-no-new-old-SMWMR}. 
\renewcommand{\toto}{theorem:evAtomocity}
\end{proofT}

\end{document}